\documentclass[aps,prx,superscriptaddress,twocolumn,floats]{revtex4-2}

\usepackage{amsmath}
\usepackage{ulem}
\usepackage{amsfonts}
\usepackage{graphicx}
\usepackage[dvipsnames]{xcolor}
\usepackage{graphicx,color}
\usepackage{tikz}

\newcommand{\nn}{\nonumber}
\newcommand{\bs}{\boldsymbol}
\usepackage[colorlinks,linkcolor=red,citecolor=blue,filecolor=magenta,
urlcolor=cyan]{hyperref}

\begin{document}
\title{Yu-Shiba-Rusinov qubit}
\author{Archana Mishra}
\email{mishra@MagTop.ifpan.edu.pl}
\affiliation{International Research Centre MagTop, Institute of Physics, Polish Academy of Sciences,
Aleja Lotnikow 32/46, PL-02668 Warsaw, Poland}
\author{Pascal Simon}
\email{pascal.simon@u-psud.fr}
\affiliation{Universit\'e Paris-Saclay, CNRS, Laboratoire de Physiques des Solides, 91405, Orsay, France}
\author{Timo Hyart}
\email{Timo.Hyart@MagTop.ifpan.edu.pl}
\affiliation{International Research Centre MagTop, Institute of Physics, Polish Academy of Sciences,
Aleja Lotnikow 32/46, PL-02668 Warsaw, Poland}
\affiliation{Department of Applied Physics, Aalto University, 00076 Aalto, Espoo, Finland}
\author{Mircea Trif}
\email{mtrif@MagTop.ifpan.edu.pl}
\affiliation{International Research Centre MagTop, Institute of Physics, Polish Academy of Sciences,
Aleja Lotnikow 32/46, PL-02668 Warsaw, Poland}
\date{\today}

\begin{abstract}

Magnetic impurities in $s$-wave superconductors lead to spin-polarized Yu-Shiba-Rusinov (YSR) in-gap states. Chains of magnetic impurities offer one of the most viable routes for the realization of Majorana bound states which  hold a promise for topological quantum computing. However, this ambitious goal looks distant since no quantum coherent degrees of freedom have yet been identified in these systems. To fill this gap we propose an effective two-level system, a YSR qubit, stemming from two nearby impurities. Using a time-dependent wave-function approach, we derive an effective Hamiltonian describing the YSR qubit evolution as a function of distance between the impurity spins, their relative orientations, and their dynamics. We show that the YSR qubit can be controlled and read out using state-of-the-art experimental techniques for manipulation of the spins. Finally, we address the effect of spin noise on the coherence properties of the YSR qubit, and show a robust behaviour for a wide range of experimentally relevant parameters. Looking forward, the YSR qubit could facilitate the implementation of a universal set of quantum gates in hybrid systems where they are coupled to topological Majorana qubits.
\end{abstract}
\maketitle
%%%%%%%%%%%%%%%%%%%%%%%%%%%%%%%%%%%%%%%%%%%%%%%%%%%%%%%%%%%%%%%%%%%%%%%%%%%%%%%%%%%%%%%%%%%%%%%%%%%%%%%%%%%%%%%%
\section{Introduction}
\label{sec1}

The goal to build a fault tolerant quantum computer has allowed to deepen the understanding of the quantum realm in a plethora of systems, as well as to an advancement in developing novel quantum technologies. Trapped ions, semiconductor quantum dots, superconducting circuits and hybrid semiconductor-superconductor platforms are some of the examples which have played crucial role in developing the field of quantum computing \cite{vion2002,yu2002, koch2007,schreier2008, michler2009, ladd2010,  barends2014,zhou2016,bruzewicz2019, arute2019, aguado2020, kjaergaard2020,zhong2020, pan2021}. 

While superconducting-circuit based qubits have been at the forefront of the immense recent progress,  proposals that utilize the low-energy bound states in superconductors, i.~e.~the Andreev levels, have been also under intense scrutiny for quantum computing \cite{chtchelkatchev2003,zazunov2003,janvier2015,park2017,tosi2019,hays2021,cerrillo2021}. The reasons are two-fold: ($i$) the dimensions of Andreev states-based qubits ($\sim\mu$m) are typically much smaller than the sizes of the conventional superconducting qubits ($\sim$mm), which facilitates designing quantum registers with higher qubit densities and ($ii$) they constitute the building blocks of topological quantum computers based on Majorana zero modes, which have experienced significant theoretical and experimental research efforts  \cite{Alicea2011, Hyart13, Aasen_milestones_2016, Karzig17, Beenakker20, flensberg2021engineered}. 
    
\begin{figure}[t]
\centering
\includegraphics[width=0.85\linewidth,trim=100mm 00mm 177cm 00mm,clip]{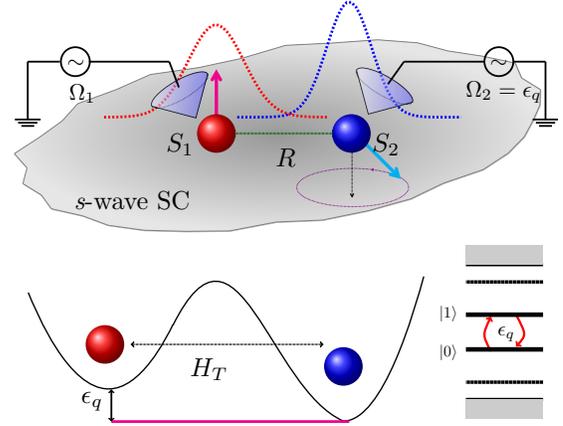}
\caption{The YSR qubit: Two classical spins, target (red) and test (blue), are placed on top of a 2D $s$-wave superconductor at a distance $R$ inducing a double-well potential that accommodates two in-gap YSR states for a given parity. The odd parity states $|0\rangle$ and $|1\rangle$ define the YSR qubit states. The asymmetry of the potential stems from the slightly different coupling parameters at the two sites. The hybridization of the two YSR states is quantified by a tunneling Hamiltonian $H_T$. Driving the test spin effectively tunes the potential bias and, when at resonance with the qubit splitting $\Omega_2=\epsilon_q$, it allows for coherent rotations of the qubit. The target spin  is interrogated off-resonantly at a frequency $\Omega_1\neq\epsilon_q$ for quantum non-demolition detection of the YSR qubit state. Alternatively, the qubit can be also operated with just a single tip.}
\label{fig:1}
\end{figure} 
    
Magnetic impurities in superconductors lead to localised Yu-Shiba-Rusinov (YSR) in-gap Andreev states \cite{yu1965,shiba1968,rusinov1969,bauriedl1981,menard2015,heinrich2018, wang2021}, 
with  chains and lattices of impurities being viable setups to realize topological superconductors hosting the Majorana modes \cite{choy2011,nakosai2013,nadj2013,braunecker2013,klinovaja2013,vazifeh2013,pientka2013, Nad14,pientka2014,poyhonen2014,heimes2014,reis2014,weststrom2015,peng2015,rontynen2015,braunecker2015, zhang2016,hoffman2016,kimme2016,neupert2016, andolina2017,schneider2021,schneider2021a,kezilebieke2020,poyhonen2014}. The advantage of these implementations is rooted in the ability to pattern superconducting surfaces with magnetic impurities, and possibly engineer (topological) quantum processors in a controlled fashion. Moreover, through the use of scanning tunneling microscopy (STM) techniques, they can be interrogated locally, with high spatial resolution. A drawback, however, is that the system parameters are hard to tune making it difficult to control the topological regime of the system, or to manipulate the emerging Majorana modes. Several solutions have been put forward, among which are exploiting the dynamics of the magnetic impurities \cite{kaladzhyan2016,kaladzhyan2017}, driving the YSR states with microwave fields \cite{AkkaravarawongPRR19}, varying the orientation of external magnetic fields \cite{JianNatComm16, kreisel2021tunable}, or tuning the Josephson effect through a superconducting tip coupled to the YSR states \cite{karan2021superconducting}.

The realization of the Majorana-based topological quantum computer in Shiba chains looks distant as no experimental evidence of quantum degrees of freedom yet exist in these systems. For this purpose it would be necessary to experimentally demonstrate that it is possible to coherently manipulate the Majorana qubits before they decohere.
 In this paper, we show that the minimal system for the demonstration of the quantumness of these systems is a new type of superconducting qubit, the YSR qubit, stemming from two nearby impurities.  We demonstrate that the dynamics of the magnetic impurities can be used for controlling the quantum state of the YSR qubit and we uncover the requirements for experimentally observing Rabi oscillations in this system.
The precession of the magnetic impurities also leads to a feedback torque acting on the impurities due to the YSR states \cite{mishra2020}, and we show that this effect can be utilized for the read out of the YSR qubit states. We also address the effect of the spin noises on the coherence properties of the YSR qubit, and show a robust behaviour for a wide range of experimentally relevant parameters. Our proposal is feasible with state-of-the-art experimental techniques, because controlled coupling of YSR states in impurity dimers have already been experimentally demonstrated \cite{kezilebieke2018,ruby2018,ding2021,beck2021}  and the manipulation of the impurity spins is possible through the STM electron spin resonance (STM-ESR) techniques \cite{balatsky2012,natterer2017,willke2018,yang2019,van2021}. Finally, we discuss the possibilities to utilize the YSR qubits in hybrid systems where they are coupled to  Majorana qubits.

The paper is organized as follows.  In Sec.~\ref{sec2}  we introduce the model Hamiltonian describing the dynamical spin dimer. Using a time-dependent wave-function approach, in Sec.~\ref{sec3} we derive the effective YSR qubit Hamiltonian in the presence of the precessing spins. In Sec.~\ref{sec4}, we discuss how to implement coherent Rabi oscillations of the YSR qubit and provide a specific manipulation protocol. Then, in Sec.~\ref{sec5} we demonstrate that spin dynamics can be utilized for the read-out of the YSR qubit. In Sec.~\ref{sec6} we introduce a hybrid YSR qubit $-$ Majorana (topological) qubit that can be operated to achieve a universal set of quantum gates. We conclude with a discussion in Sec.~\ref{sec7}.

%%%%%%%%%%%%%%%%%%%%%%%%%%%%%%%%%%%%%%%%%%%%%%%%%%%%%%%%%%%%%%%%%%%%%%%%%%%%%%%%%%%%%%%%%%%%%%%%%%%%%%%%%%
   
\section{Model Hamiltonian}
\label{sec2}    
 The time-dependent Bogolioubov de Gennes (BdG) Hamiltonian describing the spin dimer system in Fig.~\ref{fig:1} can be written in the Nambu basis
 
  $\Psi({\bs r})=[c_{\uparrow}({\bs r}), c_{\downarrow}({\bs r}),  c^\dagger_{\downarrow}({\bs r}), -c^\dagger_{\uparrow}({\bs r})]^T$ as 
\begin{align}
\label{eq:1}
 H_{\rm BdG}(t)&=H_0+\sum_{j=1,2}V_j(t)\delta({\bs r-\bs R}_j)\,,\\
H_0&=\epsilon_p\tau_z+\Delta\tau_x\,,\nn\\
V_j(t)&=J_j{\bs S_j(t)}\cdot{\bs\sigma}\,,\nn
\end{align}
where  $H_0$ is  the   superconductor Hamiltonian and $V_j(t)$ describes the coupling of electrons to the classical spins ${\bs S}_j(t)=S[\sin\theta_j(t)\cos\phi_j(t),\sin\theta_j(t)\sin\phi_j(t),\cos\theta_j(t)]$  with time-dependent polar $\theta_j(t)$ and azimuthal $\phi_j(t)$  angles ($j=1,2$). Here,  $\bs R_j= 0 ({\bs R})$ is the position of the spin $j=1$ ($j=2$),  $J_j$  are the coupling strengths,  ${\bs\sigma}=(\sigma_x,\sigma_y,\sigma_z)$ and ${\bs\tau}=(\tau_x,\tau_y,\tau_z)$ are the Pauli matrices in the spin and particle-hole spaces, $\Delta$ is the superconducting order parameter and $\epsilon_p=p^2/2m-\mu$ is the kinetic energy of the electrons with effective mass $m$, momentum $p$, and chemical potential $\mu$.  For simplicity,  we neglect the scalar potentials \cite{schneider2021} generated by the magnetic impurities as they do not affect directly the dynamics. The target spin $\bs S_1$ and test spin $\bs S_2$ can be addressed and driven individually  through STM-ESR, and they are used for read out and manipulation, respectively.
     
Before proceeding with a detailed description of the dynamics, let us provide some physical insights to the spin dimer in Fig.~\ref{fig:1} based on the recent findings in Ref.~\onlinecite{mishra2020} concerning the dynamics of a single magnetic impurity in an $s$-wave SC. We first note that  for the static case, the Shiba energy is given by $E_S=\Delta (1-\alpha^2)/(1+\alpha^2)$, with $\alpha=\pi\nu_0JS$ and $\nu_0$ being the density of states at the Fermi level in the normal state. For a spin precessing with frequency $\Omega\ll\Delta$ (adiabatic limit) at an angle $\theta$ around the $z$ axis, the effective Shiba energy was found to be $E_S(\Omega)\simeq  E_{S}-(\Omega/2)\cos{\theta}$, i.~e.~it is shifted by the Berry phase contribution. Moreover, this dynamical YSR state was found to act back on the classical spin via a {\it universal} torque ${\bs \tau}_S(t)=-(n_S-1/2)F_S\,\dot{\bs n}(t)$, where ${\bs n}(t)={\bs S}(t)/S$, $n_S$ is the YSR state occupation number and $F_S$ is the radial Berry curvature. In the absence of spin-orbit interaction, $F_S=1/2$. This torque modifies the bare resonance frequency $\Omega_0$ of the classical spin as  $\delta\Omega/\Omega_0\approx(1/S)(n_S-1/2)F_S$, being a direct measurement of the occupation $n_S$.  The energy shift (on the Shiba side) and the frequency shift (on the classical spin side) are at the core of our proposal depicted in Fig.~\ref{fig:1}: the former allows to control the bias of the double well potential by driving one of the spins, analogously to tuning the voltage-bias in double quantum dots \cite{HansonRMP2006}, while the latter facilitates extracting the occupation of the in-gap states, in analogy to quantum non-demolition qubit readouts in cavity quantum electrodynamics setups \cite{BurkardNPR20}. The hybridization between the two YSR states will modify the single-impurity findings, and in the following we proceed to describe in detail the dynamical YSR dimer system. 

%%%%%%%%%%%%%%%%%%%%%%%%%%%%%%%%%%%%%%%%%%%%%%%%%%%%%%%%%%%%%%%%%%%%%%%%%%%%%%%%%%%%%%%%%%%%%%
\section{Effective qubit Hamiltonian}
\label{sec3}

Next we derive the low-energy Hamiltonian describing the in-gap ``molecular''   YSR states stemming from the dynamical spin dimer using a time-dependent wave function approach which will allows us to identify the effective two-level system defining the YSR qubit. The system dynamics is described by the time-dependent BdG equation $i\partial_t\psi({\bs r},t)=H_{BdG}(t)\psi({\bs r},t)$, where $\psi({\bs r},t)=[u_{\uparrow}({\bs r},t), u_{\downarrow}({\bs r},t),  v_{\downarrow}({\bs r},t), -v_{\uparrow}({\bs r},t)]^T$ is the BdG wave-function. It is instructive to switch to the Fourier space $\psi({\bs r},t)= \frac{1}{L^d}\sum_{\bs k}e^{-i{\bs k}\cdot{\bs r}}\psi({\bs k},t)$ ($d$ is the dimension of the system), which in turn allows us to write
\begin{align}
    \left(i\frac{\partial}{\partial t}-H_0({\bs k})\right)\psi({\bs k},t)=\sum_{j=1,2} V_j(t)\psi({\bs r}_j,t)e^{i{\bs k}\cdot{\bs r}_j}\,.
\end{align}
Assuming that the Shiba energies are close to the Fermi level (deep Shiba limit $\alpha_{1,2}\approx 1$) and adiabatic dynamics of the classical spins on the scale of  $T_\Delta=\hbar/\Delta$, we can follow the approach described in Ref.~\cite{pientka2013,kaladzhyan2017} to derive an effective time-dependent  $8\times 8$ Schrodinger equation that describes the dimer  $i\partial_t\tilde\psi_i(t)=H_{ij}(t)\tilde\psi_j(t)$. Here,  $\tilde\psi_{1(2)}(t)=\alpha_{1(2)}\psi(0(\bs R),t)$ is a $4$-component spinor at position $r=0$ ($r=R$) in the Nambu and spin space, while  
the diagonal elements $H_{11(22)}(t)\equiv H_{1(2)}(t)$ describe the interaction of SC with the spins at the positions ${\bs r}=0 ({\bs R})$
\begin{align}
     H_{i}(t)&\approx-\Delta\left(\frac{{\bs n}_i\cdot{\bs \sigma}}{\alpha_i}+\tau_x\right)+({\bs n}_i\times\dot{{\bs n}}_i)\cdot{\bs \sigma}\label{on_site}\,,
\end{align}
and $H_{12}(t)=H_{T}(t)$ represents the tunnelling between the YSR states at different impurities
\begin{align}
    H_{T}(t)&=-\Delta({\bs n}_1\cdot{\bs \sigma})({\bs n}_2\cdot{\bs \sigma})[ \tilde I_0(R) \tau_x+\tilde I_1(R)\tau_z]\,.
    \label{tunnel}
\end{align}
Here, ${\bs n}_i(t)={\bs S}_i(t)/S_i$ and $ \tilde I_{0,1}(R)$ are evaluated from the overlap integrals for two impurities separated by a distance $R$ in the superconductor (Appendix \ref{app:1}). %\cite{SM}. 
We point out that the time-dependence of the classical spins generates a Berry-phase contribution (the second term in Eq.~\eqref{on_site}) that cannot be captured by only forging the effective static theory time-dependent. As shown later, while this term does not affect the qubit Hamiltonian, it does change drastically the spin expectation values at each impurity, in particular the contributions perpendicular to the instantaneous classical spin directions (which are responsible to the torques acting on the latter). This is one of the instances when an effective static theory does not suffice to describe the low-energy sector dynamics. 
   \begin{figure}[t]
    \centering
    \includegraphics[width=1.0\linewidth,trim=135mm 0mm 00mm 0mm,clip]{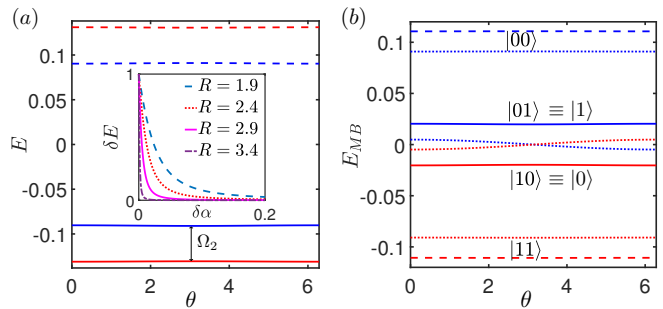}
    \caption{($a$) Single particle energy spectrum $E$ for the in-gap states as a function of $\theta$ for $\alpha_1=1.15$, $\alpha_2=1.1$ at $R=2.9$. Here $\Omega_2$ is the classical spin precession frequency that matches the qubit splitting. The inset shows the energy width $\delta E=\left[\epsilon_q(0)-\epsilon_q(\pi)\right]/\left[\epsilon_q(0)+\epsilon_q(\pi)\right]$ as a function of $\delta\alpha=\alpha_1-\alpha_2$, and for several values of $R$. ($b$) The many-body energy spectrum $E_{MB}$ for $\alpha_1=1.15$, $\alpha_2=1.1$ and $R=2.9$ in the absence of Coulomb interactions. The dotted lines represent the many-body energy spectrum for $\alpha_1=\alpha_2=1.1$,  showing a crossing at $\theta=\pi$. Here, $\{|10\rangle, |01\rangle\}$ and  $\{|00\rangle,~|11\rangle\}$ label the odd  and even parity states, respectively, with the YSR qubit being encoded in the former. In both plots we have used $k_F=13.55$. }
       \label{fig:2}
             \end{figure}

Let us first consider the dimer in the absence of dynamics. Projecting the above $8\times8$ Hamiltonian blocks, $H_i$ and $H_{T}$, onto the low energy sector results in an effective $4\times4$ Hamiltonian describing the in-gap states \cite{pientka2013,kaladzhyan2017}, (see Appendix \ref{app:1} for  details). Assuming $\theta_1=0$ (i.~e., the first spin defines the $z$-axis) the in-gap energy spectrum  of the $4\times4$ Hamiltonian becomes  $|E_{1,2}|=(B\pm C)/(2\alpha_1\alpha_2)$, where 
\begin{align}
 B&=\sqrt{(2\alpha_1\alpha_2-\alpha_1-\alpha_2)^2+\left(t_h\cos\left(k_FR+\frac{\pi}{4}\right)\sin\frac{\theta}{2}\right)^2}\nonumber\,,\\
 C&=\sqrt{(\alpha_1-\alpha_2)^2+\left(t_h\sin\left(k_FR+\frac{\pi}{4}\right)\cos\frac{\theta}{2}\right)^2}\,,
\end{align}
with $t_h=4\alpha_1\alpha_2e^{-R}/\sqrt{2\pi k_FR}$ quantifying the tunneling strength, $k_F$ being the Fermi momentum. %and $\bar{\alpha}_{1,2}=1-\alpha_{1,2}$. 
Note that all energies are expressed in terms of $\Delta=1$, while all  lengths in terms of 
SC coherence length $\xi=v_F/\Delta$, with $v_F$ being the Fermi velocity. In Fig.~\ref{fig:2}a, we show the corresponding energy spectrum as a function of $\theta$. The inset of Fig.~\ref{fig:2}a depicts the relative maximum deviation in the energy difference between the lowest two energy states as a function of $\delta\alpha=\alpha_1-\alpha_2$ for various separation distances $R$. From these plots, we can infer that $(i)$ even for moderate values of $\delta\alpha$ the dependence of the energies $E_i$ on $\theta$ is negligible and ($ii$) generally these energies are not equidistant. We can then encode the YSR qubit in the two lowest energy states defined by the $\{-E_1,-E_2\}$ which, for a wide range of parameters, are also well separated from the excited pair  $\{E_1,E_2\}$. The qubit Hamiltonian can be  written as $H_q=(\epsilon_q/2)\Sigma_z$, where $\epsilon_q\equiv\epsilon_q(\theta)=C/(\alpha_1\alpha_2)$ is the qubit splitting, and $\Sigma_z$ is the $z$ component of the Pauli matrix acting on the states defined by $\{-E_1,-E_2\}$.  

It is useful to describe the YSR qubit using a many-body states $|n_1n_2\rangle$, where  $n_{1,2}=0,1$ are the occupancy of the single quasi particle states. Specifically, the pair of states  $\{|00\rangle,|11\rangle\}$ ($\{|01\rangle, |10\rangle\}$) span the even (odd) parity many-body states with energies  $\pm|E_1+E_2|/2$ ($\pm|E_1-E_2|/2$). Note that within the BdG description the two parity sectors are decoupled, and the YSR qubit defined above acts within the odd-parity states. This choice for the YSR qubit is further justified by its insensitivity to the Coulomb interaction effects that are present for double occupancy (even parity). The many-body energy spectrum depicting the odd and even parity states is shown in Fig.~\ref{fig:2}b. While the ground state corresponds to the even parity state $|11\rangle$ for the chosen parameters, the odd parity sector can be selected by tuning the offset charge with a gate voltage in the case of a finite superconducting island with a sufficiently large charging energy. Alternatively, one can utilize the spin dynamics for the initialization of the system to the odd parity state. 

The many-body picture also allows us to gain further insight on the origin of the qubit states, which is determined by max$(\delta\alpha, t_h)$. For $\delta\alpha\gg t_h$, the qubit states stem from the two individual YSR states formed under each of the impurities, while in the opposite regime $\delta\alpha\ll t_h$, they correspond to the symmetric and anti-symmetric superposition of the individual YSR states, being  dictated by the tunneling. The first scenario is more advantageous as the qubit energies become insensitive to $\theta$, as depicted in Fig.~\ref{fig:2}, rendering it more robust against fluctuations.

Having defined the YSR qubit, we can now reinstate the dynamics of the classical spins which we will exploit for the manipulation and read out of the qubit states. Without loss of generality, in the following we assume that only one spin  precesses. Projecting the $4\times4$ time-dependent Hamiltonian  onto the YSR qubit subspace, we  obtain the following qubit Hamiltonian (Appendix \ref{app:2}):
\begin{align}
    H_q(t)=\frac{\epsilon_q}{2}\Sigma_z+{\bs \beta}(t)\cdot{\bs \Sigma}\,,
\label{ham_q}    
\end{align}
where
 \begin{align}
    \beta_{x}(t)&=t_h\frac{\sin(k_FR+\pi/4)\sin\theta\sin(\theta/2)}{4\alpha_1\alpha_2\epsilon_q}\,\dot\phi\nn \,,\\
    \beta_{y}(t)&=t_h\frac{(\alpha_1-\alpha_2)\sin(k_FR+\pi/4)\sin(\theta/2)}{4(\alpha_1\alpha_2)^2\epsilon_q^2}\,\dot\theta\,,\nn\\
    \beta_{z}(t)&=\frac{(\alpha_2-\alpha_1)\sin^2(\theta/2)}{2\alpha_1\alpha_2\epsilon_q}\,\dot\phi\,.
    \label{eq:7}
\end{align}
Eqs.~\eqref{ham_q} and \eqref{eq:7} establish the imprints of the classical spin dynamics on the effective YSR qubit Hamiltonian and represent one of our main findings. Above, we disregard the terms that act as identity in the qubit space.  The first two terms in Eq.~\eqref{eq:7} induce transitions between the qubit states, while the last term allows to dynamically control the qubit splitting $\epsilon_q\rightarrow\epsilon_q+2\beta_z(t)$. For $\delta\alpha\gg t_h$, $\beta_{x,y}(t)\propto t_h/\delta\alpha$, while $\beta_z(t)\approx \sin^2(\theta/2)\dot{\phi}/2$ is independent of any of the microscopic parameters. 

%%%%%%%%%%%%%%%%%%%%%%%%%%%%%%%%%%%%%%%%%%%%%%%%%%%%
\section{YSR qubit manipulation}
\label{sec4}

     \begin{figure}[t]
    \centering
    \includegraphics[width=\linewidth,trim=100mm 0mm 00mm 0mm,clip]{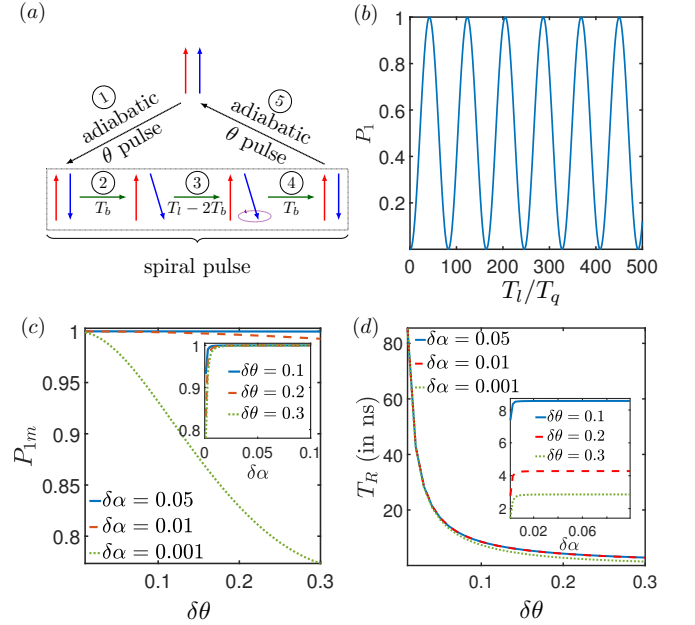}
    \caption{($a$) The pulse sequences proposal for coherent manipulation of the YSR qubit: the right spin (blue) is first rotated adiabatically from $\theta=0$ to $\theta=\pi$  in time $T_a$, then driven into resonant precession with the qubit for a time $T_l$, after which it is rotated back adiabatically to its original orientation. The qubit can also be fully operated in the anti-parallel configuration. ($b$) The Rabi oscillations encoded in the probability of state $|1\rangle$ being occupied, $P_1(T_l)=|\langle1|\psi(T_l)\rangle|^2$, as a function of the pulse time $T_l$ for $T_a=6T_q$ and  $T_b=0.6T_q$. The other parameters are $\delta\theta=0.1$, $\alpha_1=1.15$,  $\alpha_2=1.1$, $R=2.9$, so that $t_h/\delta\alpha=0.34$. ($c$) The Rabi oscillations amplitude, $P_{1m}$, as a function of the angle deviation $\delta\theta$ (main) and $\delta\alpha$ (inset). Here, $t_h/\delta\alpha=17, 1.7, 0.34$ for $\delta\alpha=0.001,0.01,0.05$, consistent with $P_{1m}\sim1$ for $t_h\ll\delta\alpha$ on a wide range of $\delta\theta$ values. ($d$) The Rabi oscillation period, $T_R$, as a function of $\delta\theta$ (main) and $\delta\alpha$ (inset) for different values of $\delta\alpha$ and $\delta\theta$, respectively.  The same conclusions as in $(c)$ apply. All plots have been obtained using $k_F=13.55$.} 
    \label{fig:4}
\end{figure} 
The YSR qubit can be manipulated by utilizing the second term in Eq.~\ref{ham_q}. The pulse sequence for introducing Rabi oscillations is shown schematically in Fig.~\ref{fig:4}a. The logical states of the qubit are defined in a parallel classical spins alignment, and the resonant oscillations between the states of the qubit are induced in the anti-parallel configuration. Before describing the details of the sequence, let us underline the physical reasons for this choice. The $\beta_i(t)$ terms in Eq.~(\ref{eq:7}) are much weaker for deviations $\delta\theta$ around $\theta=0$ ($\beta_{x,z}\propto(\delta\theta)^2$ and $\beta_{y}\propto\delta\theta$) than when the same deviations occur in proximity of $\theta=\pi$ ($\beta_{x}\propto\delta\theta$ and $\beta_{y,z}\propto$ constant), which makes them rather inefficient in the parallel configuration. In the idle phase, on the other hand, this is beneficial since the qubit will be more robust against random fluctuations in the angles $\theta$ and $\phi$ (discussed below). Nevertheless, the qubit can  also be operated fully in the anti-parallel geometry, at the expense of shorter coherence times.   

Let $|0\rangle$ and $|1\rangle$ be the eigenstates of the static qubit Hamiltonian at $\theta=0$, and assume the qubit is initialized in state $|0\rangle$ at time $t=0$. Then, at time $t$, the qubit  state becomes
$|\psi(t)\rangle=U_q(t,0)|0\rangle$ where the evolution operator is $U_q(t,0)=\mathcal{T}e^{-\frac{i}{\hbar}\int^{t}_{0}d t'H_q(t')}$ with $\mathcal{T}$ being the time-ordering operator. In step \raisebox{.5pt}{\textcircled{\raisebox{-.9pt} {1}}} of the protocol in Fig.~\ref{fig:4}a,  the right classical spin is rotated from parallel to the anti-parallel configuration via a pulse $\theta(t)=\pi\tanh(2\pi t/T_a)$ where $T_a$ is the pulse length, and the evolution operator is $U_{q,1}\equiv U_q(T_a,0)$ \cite{mayergoyz2009}.  The amplitude of the Rabi oscillations is largest if the the qubit remains in state $|0\rangle$ during this pulse. Thus, ideal results are obtained if $0-\pi$ transition is adiabatic, i.~e.~$T_a\gg T_q$  with $T_q=\hbar/\epsilon_q$, but almost ideal Rabi oscillations can be achieved also for fast pulses (Appendix \ref{app:5}). In the second part of the sequence, the classical spin is driven into circular precession around the $z$ axis so that the precession frequency $\Omega_2$ is in resonance with the qubit splitting   $\Omega_2=\epsilon_q$. Consequently, the qubit undergoes coherent Rabi oscillations, and the evolution is described by $U_{q,2}\equiv U_q(T_a+T_l,T_a)$. In our calculations we use a spiral pulse $\phi(t)=\Omega_2 t$ and 
\begin{equation}
\theta(t)=\pi-\delta\theta\left(\tanh\frac{t}{T_b}-\tanh\frac{t-T_l}{T_b}-1\right)\,,
\label{spiral}
\end{equation}
which first stabilizes the precession of the spin to a cone angle $\theta=\pi-\delta\theta$ in a time $T_b$ (step \raisebox{.5pt}{\textcircled{\raisebox{-.9pt} {2}}}), then causes a precession of the spin for a duration $T_l-2T_b$ (step \raisebox{.5pt}{\textcircled{\raisebox{-.9pt} {3}}}) and finally restores the classical spin back to $\theta=\pi$ in time $T_b$ (step \raisebox{.5pt}{\textcircled{\raisebox{-.9pt} {4}}}). Assuming $T_b\ll T_q$ implies that the evolution induced by $\dot{\phi}(t)$ during the ramping periods is practically frozen and we can write $U_{q,2}\approx U^{-1}_{q,2g}U_{q,2r}U_{q,2g}$, where $U_{q,2g}$ and $U_{q,2r}$ correspond to the evolution from $\pi$ to $\pi-\delta\theta$ at $\dot{\phi}=0$ and the evolution induced by the circular precession at fixed $\delta\theta$ ($\dot{\theta}=0$) during the time $T_l-2T_b$, respectively. Finally,  the classical spin is rotated back to the parallel configuration using $U_{q,3}\equiv U^{-1}_{q,1}$ shown by step \raisebox{.5pt}{\textcircled{\raisebox{-.9pt} {5}}} in Fig.\ref{fig:4}a. 

% \noteMT
The amplitude and the period of the Rabi oscillations can be determined by calculating how the  probability for the qubit to be in state $|1\rangle$ after a pulse,  $P_{1}(T_l)=|\langle1|\psi(T_l)\rangle|^2$ with $|\psi(T_l)\rangle=U_q(T_l)|0\rangle$, depends on the precession time $T_l$. We have implemented numerically the evolution operator pertaining to $U_q(t)$, and in Fig.~\ref{fig:4}b we plot $P_{1}(T_l)$ showing the Rabi oscillations of the qubit for the parameters $\delta\alpha=0.05,~\delta\theta=0.1,~R=2.9$. Increasing the precession angle $\delta\theta$ increases the Rabi oscillations frequency as  $\Omega_R\propto t_h\delta\theta$, but in turn reduces their amplitude, as depicted in Figs.~\ref{fig:4}c,d. The latter is a consequence of the transformation $U_{q,2g}$ which generates a finite weight $c_1\propto(t_h/\delta\alpha)\delta\theta$ on the state $|1\rangle$ for $t_h\ll\delta\alpha$. Therefore, for a given $\Omega_R$, the requirement for $P_{1m}\equiv{\rm max}[P_1(T_l)]\approx 1$ is $\Omega_R\ll\epsilon_q$ which, coincidentally, is similar to the adiabaticity condition in the first part of the protocol. Moreover, in the limit $T_b\ll T_q$ the transformation $U_{q,2g}$ is purely geometrical (Appendix \ref{app:5}), and thus independent on the details of the pulse that tilts the classical spin away from the $z$ axis by an angle $\delta\theta$. As stressed above, the manipulation can be fully performed in the anti-parallel configuration, in which case $U_q(T_l,0)\equiv U_{q,2}$. Both the parallel and anti-parallel configurations have been observed experimentally, their realization depending on the specific implementation and the distance between the impurities \cite{choiPRL18,ding2021}. 
% \noteMT{Below (until the end of this section) the text is changed.}

To give some estimates for the time scale of the Rabi oscillations, let us assume $\delta \theta=0.1,~R=2.9$ and $\delta\alpha=0.05$. These rather conservative parameter values result in Rabi oscillation period  $T_R\approx\,8.5$ ns, which is comparable to the Rabi times observed  in implementations of the  Andreev qubits \cite{janvier2015}. For the YSR qubit to be useful, the Rabi time should be much shorter than the time scales over which it looses its coherence, namely the relaxation ($T_1$) and pure dephasing ($T_\phi$) times, which  to the best of our knowledge, are largely unknown for the YSR states. Nevertheless, we can readily identify several possible sources of decoherence: ($i$) quasiparticle poisoning \cite{Dima18, mannila2021superconductor,Dima21},  ($ii$) thermal fluctuations in the magnetic moments (magnons) that define the YSR states, and ($iii$) phonon or photon coupling to the Shiba electrons \cite{ruby2015}. Decoherence induced by non-equilibrium quasiparticle poisoning is highly specific to the system and thus it is difficult to provide precise scalings and estimates. Recent studies, both experimental and theoretical, show that the relaxation times pertaining to this mechanism can range from milliseconds to even seconds \cite{Dima18, mannila2021superconductor,Dima21}. The general consensus is that their effect can be minimized by improving the samples, and it can be accounted for by a phenomenological line-width of the isolated YSR states,  which in-principle can be extracted from STM-ESR measurements in the limit of weak tunnel coupling \cite{huang2020, huang2021}. The last two mechanisms, on the other hand, have not been discussed in the literature for the YSR molecule. In the following, we give a short account of the magnons-induced decoherence, while the details of the phonon (and photon) mechanism is described in Appendix \ref{app:6} and \ref{app:7}. The Hamiltonian describing the coupling of the qubit to the magnetization fluctuations of spins $k=1,2$ reads:
\begin{align}
\delta H_{q}(t)&=\sum_{k=1,2}\delta {\bs n}_{k}(t)\cdot\,{\bs\chi}_{k}\cdot{\bs \Sigma}\,,
\end{align}
with the tensor ${\bs \chi}_{k}\equiv[\chi_k^{\mu\nu}]$ quantifying the coupling of the two orthogonal fluctuations $\mu=1,2$ ($\delta{\bs n}_{k}(t)\perp{\bs n}_k$) of each classical spin $k$  to the qubit Pauli matrices $\nu=x,y,z$. The elements of the tensor $\chi_k^{\mu\nu}$ can be found by projecting  $\partial_{\bs n_k}H_{BdG}$ onto the qubit basis (Appendix \ref{app:6}).  Within the Bloch-Redfield framework \cite{blum2012}, we find the following expressions for the dephasing and relaxation times, respectively:
\begin{align}
    \frac{1}{T_{\phi,m}}&=\sum_{\mu,k=1,2}|\chi_{k}^{\mu z}|^2S^k_{11}(0)\,,\\
    \frac{1}{T_{1,m}}&=\sum_{\mu,\nu,k=1,2}\chi^{\mu \sigma}_{k}\chi^{\nu \bar{\sigma}}_{k}S^k_{\mu\nu }(\sigma\epsilon_q)\,,
\end{align}
where $\chi^{\mu \sigma}_{k}=\chi^{\mu x}_{k}+i\sigma\chi^{\mu y}_{k}$ and  $S^k_{\mu\nu}(\omega)=(1/2\pi)\int dt e^{-i\omega t}\langle\delta n_{k,\mu}(t)\delta n_{k,\nu}(0)\rangle$ is the noise spectrum pertaining to the fluctuations $\delta n_{k,\mu}(t)$. Above, the $\sigma=+(-)$ terms represent the emission (absorption) rates that are related by the detailed balance condition at equilibrium. The spectrum of the fluctuations $\delta{\bs n}_{k}(t)$ is determined by the specific form of the classical spins free energy $F_S(\bs n)$ and, in order to give estimates for the above decoherence times, we consider the following form (assuming the free energies of the two spins to be identical):
\begin{align}
    F_{S}({\bs n})=-\frac{\kappa}{2}n_z^2-\gamma{\bs B}\cdot{\bs n}\,,
\end{align}
where $\kappa$ measures the crystal anisotropy (intrinsic or induced by the surface),  ${\bs B}$ is the externally applied magnetic field, and $\gamma$ the gyromagnetic ratio. Considering  $\kappa>0$, this free energy per spin is consistent with the perpendicular to the surface configurations observed in experiments.  At finite temperatures ${\bs B}\rightarrow{\bs B}+\delta{\bs B}(t)$, with $\delta{\bs B}(t)$ being the stochastic contribution whose Fourier components satisfy the fluctuation-dissipation relations $\langle \delta B_{\mu}(\omega)\delta B_{\nu}(\omega')\rangle=(\alpha_g\hbar\omega)/(\gamma^2 S)\left[\coth{(\hbar\omega/2k_BT)}-1\right]\delta(\omega+\omega')$ \cite{landau2013}, where $\alpha_g$ and $\gamma$ are the Gilbert damping and gyromagnetic coefficient, respectively. Utilizing the Landau-Liftshitz-Gilbert (LLG) equation that describes the dynamics of the classical magnets in the presence of the stochastic magnetic fields $\delta{\bs B}(t)$, we can evaluate the correlators $S_{\mu\nu}^k(\omega)$ in terms of $\langle \delta B_{\mu}(\omega)\delta B_{\nu}(\omega')\rangle$ (see Appendix \ref{app:6} for more details). For simplicity, we focus only on the static (idle) parallel and anti-parallel spin configurations, assuming a spin $S=5/2$ at each site \cite{vzitko2018,oppen2021}. 
In both cases, we find that the pure dephasing rate is zero and, furthermore for $\theta=0$, the relaxation rate is also zero, justifying quantitatively our choice for the qubit basis in the idle phase.  However, at  $\theta=\pi$ the longitudinal relaxation rate is non-zero, and  assuming $\alpha_g=0.001,~\kappa=0.1\,{\rm meV},~\alpha_1=1.15,~\alpha_2=1.1,~R=2.9$ and temperature $T_0=100\,{\rm mK}<\epsilon_q$, we obtain $T_{1,m}\approx3.5\,\mu$s \cite{hatter2017}. Comparing that to the Rabi oscillation period we estimate that the YSR qubit can undergo a large number of Rabi oscillations before it decoheres due to magnons. 

We found that both the phonon and photon couplings vanish in the anti-parallel configuration (where the YSRQ is operated), in stark contrast to the magnons which have their maximal effect. That is because both phonons and photons cannot induce spin flips, which are required for quasiparticle tunneling between the two YSR states in this configuration. In the parallel arrangement instead the phonon induced relaxation is maximal, and we evaluated it to be $T_{1,ph}\approx 5.8 \mu$s. This is a slightly longer time than the coherence time induced by the noise in the magnetic moments. However, all these sources of decoherence seem to be of similar magnitude, and they are also similar to the coherence time observed in the Andreev qubits \cite{janvier2015}. 
%%%%%%%%%%%%%%%%%%%%%%%%%%%%%%%%%%%%%%%%%%%%%%%%%% 
\section{YSR qubit read out}
\label{sec5}

The ability to measure efficiently and fast the outcome of a computation is a prerequisite for a practical qubit. Furthermore, it allows to initialize the qubit state at the beginning of the computation. Here, we show that the qubit state can be measured using STM-ESR techniques via the torques induced by the YSR states on the classical spins. In the following, we focus on the case when the measurement is performed in the parallel spin configuration  and the left spin (target) is interrogated off-resonantly with the qubit splitting as shown in Fig.~\ref{fig:1}. The former condition is considered in order to minimize the decoherence effects, while the latter allows to physically separate the manipulation and detection. 

The dynamics of the left spin, ${\bs S}_1$, is governed by the LLG equation 
           \begin{equation}
          \dot{\bs S}_1=-\gamma\bs S_1\times\bs B(t)+\bs\tau+\alpha_g \bs S_1\times \dot{\bs S}_1\,,
      \end{equation}
where ${\bs \tau}=-J_1{\bs S}_1\times\langle{\bs\sigma}\delta({\bs r})\rangle$ is the torque pertaining to the electrons in the SC that act on spin ${\bs S}_{1}$, including the YSR qubit contribution, while ${\bs B}(t)$ is the time-dependent external magnetic field utilized to drive the precession. 

\begin{figure}[t]
    \centering
    \includegraphics[width=\linewidth,trim=100mm 0mm 00mm 0mm,clip]{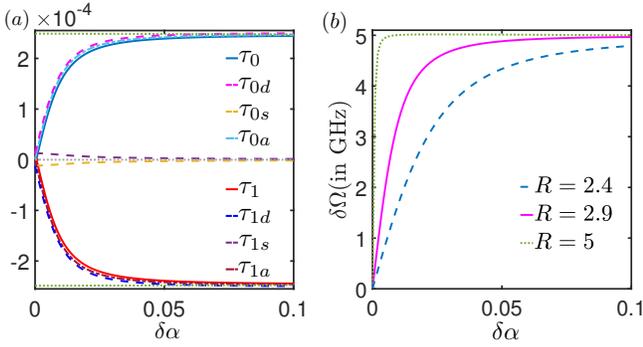}
    \caption{($a$) The  total ($\tau_\sigma$), static ($\tau_{\sigma s}$) and dynamic ($\tau_{\sigma d}$) torques as a function of $\delta\alpha$ for $\theta=0.01$, $\Omega_1/\Delta=0.1$ and $R=2.9$. The dotted lines correspond to the asymptotic behaviour of the two contributions to the torque $\tau_{\sigma s}$ and $\tau_{\sigma d}$ for $t_h\ll\delta\alpha$. 
    The  total torque evaluated analytically ($\tau_{\sigma a}$) from  Eq.~(\ref{an_torques}) shows excellent agreement with the numerical result (dot-dashed lines). ($b$) The difference of the resonance frequency in the two qubit states,  $\delta\Omega=\Omega_{r,0}-\Omega_{r,1}$, as a function of $\delta\alpha$ for $\alpha_2=1.1$, $\Omega_0=25$ GHz,  and several separations $R=2.4,~2.9$ and $5$. The saturation of $\delta\Omega$ for even moderate separations indicates robust qubit detection. All plots have been obtained using $k_F=13.55$.}
    \label{fig:3}
\end{figure}

We have employed a Green function approach that describes the $8\times8$ Hamiltonian \cite{kaladzhyan2017} to evaluate the total torque ${\bs \tau}_\sigma$ for the two YSR qubit states $\sigma=0,1$ (see Appendix \ref{app:3} for more details). Considering $\bs S_1$ to precess with frequency $\Omega_1$ in the adiabatic limit $\Omega_1\ll\Delta$, we can write 
$\bs\tau_\sigma\approx\bs\tau_{\sigma s}+{\bs \tau}_{\sigma d}$,  where the first term (${\bs \tau}_{\sigma s}$) originates from the misalignment of the two classical spins, and it describes the in-gap states contribution to the RKKY interaction, while the latter (${\bs \tau}_{\sigma d}\propto\Omega_1$) have been unravelled recently in Ref.~\onlinecite{mishra2020} and found to have a geometrical (Berry phase) origin. In Fig.~\ref{fig:3}a we show the magnitudes of the total torque $\bs\tau_{\sigma}$, as well as the two individual  contributions  $\bs\tau_{\sigma s}$ and $\bs\tau_{\sigma d}$, as a function of $\delta\alpha$ for each of the two qubit states. 
We see that the torques are  determined by the static term $\tau_{\sigma s}$ in the limit $\Omega_1|\delta\alpha|\ll t_h^2$, while in the opposite regime, $\Omega_1|\delta\alpha|\gg t_h^2$, the dynamical contribution $\tau_{\sigma d}$ dominates and, moreover, it reaches a universal value associated with an isolated impurity \cite{mishra2020}. We mention that throughout the section we have neglected the effect of the bulk states on both the static and dynamical torques. In  Ref.~\cite{YaoPRL14} it was shown that the (static) bulk contribution, which represents the conventional RKKY interaction, becomes negligible compared to that of the YSR in-gap states for separations $R\geq1$. Furthermore, a full non-equilibrium calculation for a single impurity showed that the YSR states dominate the dynamical torque in the deep Shiba adiabatic regime and, moreover, that a finite YSR   linewidth  imprints onto the magnetic impurity linewidth \cite{mishra2020}, which could be utilized to measure the coherence times of the YSR qubit.

In the limit of small cone angle precession ($\theta\sim0$), we can  linearize the LLG equation, and extract the  renormalized resonance frequency of spin ${\bs S}_1$ for each qubit state in terms of the torques as (Appendix \ref{app:4})
\begin{align}
    \Omega_{r,\sigma}=&\frac{\Omega_0-\displaystyle{\frac{\tau'_{\sigma s}}{S}}}{\displaystyle{1+\frac{\tau'_{\sigma d}}{S}}}\,,
\end{align}
where $\tau'_{\sigma d}=(1/\Omega_1)\left(\partial \tau_{\sigma d}/\partial\theta\right)|_{\theta=0}$, $\tau'_{\sigma s}=\left(\partial \tau_{\sigma s}/\partial\theta\right)|_{\theta=0}$, and  $\Omega_0$ is the bare resonance frequency. The difference  $\delta\Omega=\Omega_{r,0}-\Omega_{r,1}$ discriminates the two qubit states in STM-ESR measurements and represents one of our main findings. In Fig.~\ref{fig:3}b we plot $\delta\Omega$ as a function of $\delta\alpha$ for various  distances $R$. For separations $R$ such that $t_h^2\ll\Omega_1|\delta\alpha|$, the difference $\delta\Omega$ saturates to a constant value, which we find to be $\delta\Omega\approx8S\Omega_0/(16S^2-1)$ \cite{mishra2020}. That is because the two impurities become practically decoupled, resulting in $\tau_{\sigma s}\rightarrow0$, and only the dynamical torque from the isolated impurity contributes to the signal. We see again here that the optimal regime for operating the YSR qubit is when tunneling between the two isolated YSR states is smaller than their energy difference, in which case $\delta\Omega$ is almost invariable for wide range of system parameters.

To enrich the understanding of the above results, we present an heuristic derivation of the YSR qubit torques from basic energy considerations. In the readout regime $\dot{\theta}=0$ and $\Omega_1=\dot{\phi}\neq\epsilon_q$, so that the effective qubit splitting is $\epsilon_q^{\rm eff}=\epsilon_q+2\beta_z$ and we can neglect the $\beta_{x,y}$ terms in Eq.~(\ref{eq:7}). Then, the magnitude of the torque acting on the spin ${\bs S}_1$ by the YSR qubit in state $\sigma=0,1$ can be expressed as
$\tau_{\sigma}=(-1)^{\sigma}\frac{1}{2}\partial_\theta\epsilon_q^{\rm eff}\equiv (-1)^{\sigma}(\tau_{s}+\tau_d)$, with 
\begin{align}
    \tau_{s}&=-\frac{\displaystyle{t^2_h\sin^2\left(k_FR+\pi/4\right)\sin\theta}}{2(2\alpha_1\alpha_2)^2\epsilon_q}\,, \nonumber \\
    \tau_{d}&=-\Omega_1\frac{(\alpha_2-\alpha_1)}{4\alpha_1\alpha_2\epsilon_q}\left(\sin\theta-\frac{4\tau_{s}}{\epsilon_q}\sin^2(\theta/2)\right)\,.
    \label{an_torques}
\end{align} 
We see that for $t_h^2\ll\Omega_1|\delta\alpha|$ the dynamical torque dominates, reaching a universal value $\tau_d\approx(\Omega_1/4)\sin{\theta}$, consistent with the findings in Fig.~\ref{fig:3}a. On the other hand, for $t_h^2\gg\Omega_1|\delta\alpha|$ the torque is controlled by the static contribution $\tau_s$, and reaches the asymptotic value $\tau_s\approx -(t_h/4\alpha_1\alpha_2)\sin(k_FR+\pi/4)\sin{(\theta/2)}$ that depends strongly on the separation  $R$ between the impurities. This is again consistent with the findings in Fig.~\ref{fig:3}a. Interestingly, while for $t_h^2\ll\Omega_1|\delta\alpha|$ the torque $\tau_d$ behaves similarly around $\theta=\pi$, for $\alpha_1=\alpha_2\equiv\alpha$ we obtain $\tau_s\approx -(t_h/4\alpha^2)\sin(k_FR+\pi/4)$, which is a consequence of  the two qubit levels crossing each other: even though the spins are anti-parallel, a torque is exerted between the two, which is to be contrasted with the RKKY interactions mediated by the bulk \cite{YaoPRL14}. This can also be interpreted as a fractional spin Josephson effect that is protected by the presence of the inversion symmetry.   

To give estimates for the possible frequency shifts $\delta\Omega$, let us consider the following experimentally pertinent values: $\alpha_1=1.15,~~\alpha_2=1.1$, $R=2.9$,  $\Delta=1\,{\rm meV}\approx241$ GHz, and $S=5/2$. Interrogating the classical spin with frequencies  $\Omega_0\sim25$ GHz, results in $\delta\Omega\approx4.9$ GHz, which is well within the state-of-the-art experimental resolution \cite{yang2019}. Note that with the above parameters, the qubit splitting $\epsilon_q\approx9.5$ GHz, and thus the target spin precesses off-resonantly, which is essential for the non-invasive read-out of the qubit.  
 
%%%%%%%%%%%%%%%%%%%%%%%%%%%%%%%%%%%%%%%%%%%%%%%%%%%     
\section{YSR-Majorana hybrid qubit}
\label{sec6}

The YSR qubits described here could be utilized for quantum information tasks on their own, for example, by creating a network of weakly interacting spin dimers on top of superconductors that can be addressed individually. More importantly, they could be integrated with Majorana zero modes hosted at the ends of spin chains in superconductors and exploited for performing universal quantum computation. Indeed, the braiding statistics of the Majorana zero modes alone is not sufficient for implementing a universal set of topological gate operations necessary for quantum computation, and additional non-topological gates are needed to achieve universality. A viable way to  implement the missing $\pi/8$ phase gate is to control the couplings of the Majorana zero modes \cite{BravyiPRA05}, e.~g.~by varying the magnetic fluxes in transmon geometries \cite{Hassler_2011, Hyart13, Heck_2015}, and extremely robust geometric \cite{PhysRevX.6.031019} and distillation \cite{PhysRevA.71.022316} protocols can be utilized if sufficiently accurate control of the couplings is possible. However, it is not easy to realize suitable pulses to control the couplings of the Majorana modes in  Shiba chains, and to our knowledge there currently does not exist proposals for robust protocols to implement the $\pi/8$ gate in these systems. An alternative proposal to implement the $\pi/8$ gate is to integrate the Majorana zero modes with quantum dots-based qubits \cite{flensberg2011,LeijnsePRL11,Aasen_milestones_2016,HoffmanSWAP16}, and here we show that this idea can be transferred to the context of the Shiba chains by utilizing the YSR qubit. 

\begin{figure}[t]
    \centering
 \includegraphics[width=\linewidth]{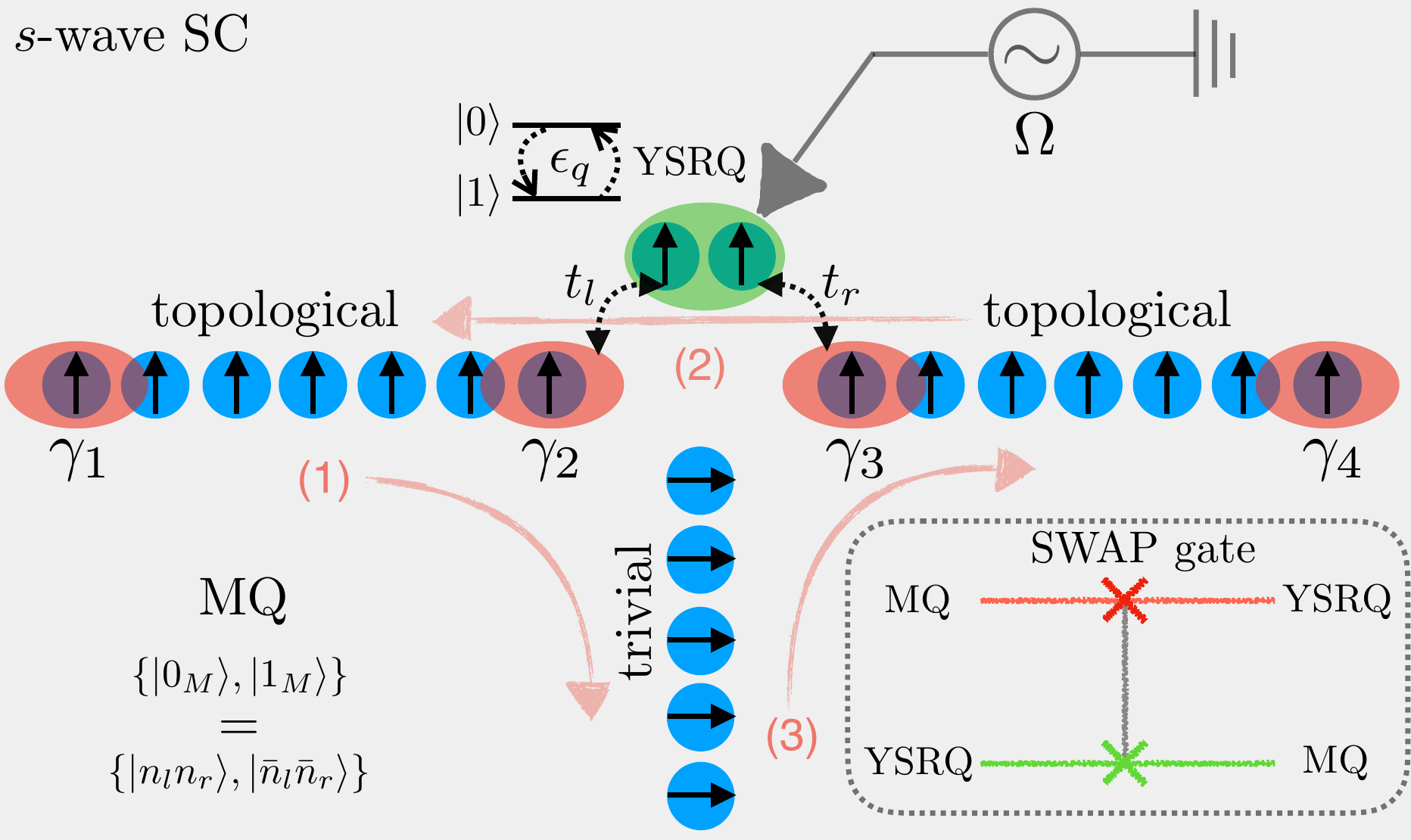}
\caption{Sketch of the hybrid YSR qubit (YSRQ) $-$ Majorana qubit (MQ). The magnetic adatoms (black arrows) deposited on the top of an $s$-wave SC form three Shiba chains in a T-junction geometry. The left ($l$) and right ($r$) chains are assumed topological, hosting Majorana zero modes $\gamma_{i}$ ($i=1,\dots,4$) at their ends (red) and are coupled via tunnelings $t_{l,r}$ to a spin dimer that accommodates a YSRQ (green), while the lower Shiba chain is non-topological. The states of the MQ are defined by the fermionic occupations $n_{l(r)}=0,1$ of the two Majorana modes on the left (right) chain, and the topologically protected Clifford gates can be implemented by braiding [steps $(1)$, $(2)$, and $(3)$] and fusing the Majorana zero modes. Driving locally the dimer rotates coherently the YSRQ (see Sec.~IV), and by combining this rotation with the SWAP gates facilitated by $t_{l,r}$ (described in detail in Ref.~\cite{HoffmanSWAP16}) the missing $\pi/8$ phase gate can be implemented. The SWAP gate can also be used for the measurement of the MQ via the readout of the YSRQ.}
       \label{fig:5}
             \end{figure}

In Fig.~\ref{fig:5} we sketch our proposal for generating a universal set of gates in a hybrid YSR qubit (YSRQ) and the Majorana qubit (MQ) system. A T-junction formed by three Shiba chains generated by magnetic impurities (or adatoms) placed on top of an $s$-wave superconductor interacts via tunnelings $t_{l,r}$  with a spin dimer that encodes a YSRQ described in the previous sections. The left and right Shiba chains are topological, hosting Majorana zero modes $\gamma_i^\dagger=\gamma_i$ ($i=1,\dots,4$), while the lower one is in the trivial regime. Two Majorana zero modes on the left (right) chain define a fermionic state, which can be occupied or empty.    Thus, the states of the MQ can be encoded either as $\{|0_l0_r\rangle,|1_l1_r\rangle\}$ (in the case of even parity) or as $\{|0_l1_r\rangle,|1_l0_r\rangle\}$ (odd parity).

The hybrid qubit can be operated as suggested in Ref.~\cite{HoffmanSWAP16}. The Clifford gates can be implemented in a topologically protected fashion by fusing Majorana zero modes and by braiding them in a T-junction shown in Fig.~\ref{fig:5} \cite{Alicea2011}. Additionally, the non-topological $\pi/8$ gate can be performed by first swapping the MQ state to YSRQ, then performing the $\pi/8$ gate on the YSRQ, and finally swapping the YSRQ state back onto the MQ.  The SWAP gate can be constructed by utilizing the general ideas presented in Ref.~\cite{HoffmanSWAP16}. The tunable couplings $t_{l,r}$ lead to an interaction Hamiltonian between the qubits
\begin{align}
    H_{\rm int}(t)=\sum_{i,j=x,y,z}J_{ij}(t)\Sigma_i\mathcal{M}_j\,,
\end{align}
where $\mathcal{M}_j$ ($j=x,y,z$) are the Pauli matrices acting on the MQ  and $J_{ij}(t)$ are  time-dependent coupling strengths originating from $t_l,t_r\neq0$. In the case of $\epsilon_q=0$ (degenerate YSRQ qubit), a specific sequence of operations for implementing the SWAP gate has been provided in Ref.~\cite{HoffmanSWAP16}. In our case $\epsilon_q=0$ can be achieved easily in the anti-parallel configuration by applying a magnetic field  $B_z$ along the $z$-axis. Indeed, for $\theta=\pi$ the tunneling vanishes, and we get that $\epsilon_q=0$ at $g\mu_BB_z=2\delta\alpha$  with $g$ and $\mu_B$ being the $g$-factor and the Bohr magneton, respectively. Moreover, since we are assuming the deep Shiba limit ($\delta\alpha\ll1$), the condition $g\mu_BB_z\ll1$ is satisfied, which  means the bulk SC remains unaffected. Alternatively, it might be possible to utilize the dynamics of the classical spin in the implementation of the SWAP gate. The SWAP gate can also be used for the initialization and measurement of the MQ via the readout of the YSRQ.

%%%%%%%%%%%%%%%%%%%%%%%%%%%%%%%%%%%%%%%%%%%%%%
\section{Discussion}
\label{sec7}    

In this work, we introduced and studied a novel type of quantum bit, the YSR qubit, that is encoded in the energy states of a spin dimer coupled to an $s$-wave superconductor.   We have demonstrated theoretically that both the coherent manipulation and the readout of the YSR qubit can be efficiently implemented by harnessing the dynamics of the spins that engenders it. Furthermore, we scrutinized the effect of the classical spins fluctuations on the coherence of the  YSR  qubit, and showed robust behaviour compared to the manipulation times. Given the ability to manipulate  magnetic adatoms on superconducting substrates with a high degree of control, the YSR qubit could be utilized together with Majorana topological qubits to facilitate performing universal quantum computation. We proposed one such hybrid implementation that is based on topological Shiba chains and a spin dimer hosting the  YSR qubit.  
            
There are several avenues for future studies. An immediate objective would be to generalize the time-dependent formalism described here to account for the spin-orbit effects originating from both the substrate \cite{beck2021} and the anisotropy of the exchange coupling between the classical spins and the superconducting electrons \cite{ding2021}. The spin-orbit coupling effects can provide the microscopic mechanism for the easy-axis anisotropy and therefore they are potentially useful for the operation of the YSR qubit, but additionally they can stabilize the ferromagnetic order in adatom chains and facilitate the realization of the Majorana modes. Thus, these effects are crucial for the operation of the hybrid qubit.     

Another important direction is to establish the  Hamiltonian of the hybrid qubit from the  microscopics in order to study the possible quantum gates and to optimally engineer the adatom deposition. Additionally, transferring coherently the information between the two types of qubits might be  beneficial for entangling MQs which are separated by a large distance. The YSRQs could be entangled for example by utilizing cavity quantum electrodynamics similarly as it has been employed in various other solid-state qubits \cite{BurkardNPR20}. While the coupling of the YSR qubit to the magnetic field of a microwave cavity should be weak ($<$ kHz), the stronger electric field component instead could couple to the qubit by affecting the tunneling $t_h$ or via the spin-orbit coupling. Alternatively, these interactions could be ignited indirectly, via the coupling of the quantum fluctuations of the classical spins to the microwave photons \cite{TabuchiPRL14,zhang2016}, which could also be used to drive them into precession. Moreover, such setups would naturally allow for the YSRQ to interact with other types of qubits, and enhance their functionality.
            
Further down the road, it would be interesting to extend the dynamical framework describing the spin dimers also to Shiba chains and 2D Shiba islands that can host Majorana end modes and  chiral Majorana edges \cite{MenardNatComm17,kezilebieke2020}, respectively. We believe that by triggering the magnetic dynamics it should be possible to both manipulate and detect the Majorana edge modes, the latter by leaving their fingerprints on the STM-ESR signals. Moreover, such approaches should be advantageous as they would allow to interrogate Shiba systems with well established methods from spintronics \cite{valenzuelaRMP15}. As catalyst for another direction of future work, we speculate  that the in-gap Shiba states, either in dimers or chains, could mediate out-of-equilibrium spin interactions in the  presence of external magnetic drives that have no counterparts in the static situations. Given the non-perturbative nature of the coupling between the electrons in the superconductor and the spins, which converts into torques as in Eq.~(\ref{an_torques}), that would require solving self-consistently the combined dynamics of the two systems. This might result in novel spin configurations that are stabilized dynamically and, on the electronic side, induce new types of (possibly dissipative) phases \cite{ghosh2021driving}.

In conclusion, the  YSR qubit proposed in this work operates well within the current experimental  capabilities, and we expect to open up new possibilities for future studies on superconducting systems patterned with spins. We hope our findings  will help create a roadmap  towards a functional MQ in these systems.

{\it Acknowledgments}$-$ We would like to thank Thore Posske for the interesting and fruitful discussions. The work is supported by the Foundation for Polish Science through the IRA Programme co-financed by EU within SG OP.

\onecolumngrid
\newpage
\appendix
\section{\label{app:1}Derivation of the low-energy Hamiltonian}
      
In this section, we show the derivation of the Hamiltonian describing the YSR qubit given by Eq.~\eqref{ham_q}. The total SC Hamiltonian written in the Nambu basis $\Psi({\bs r})=\left[c_\uparrow(\bs r),~c_\downarrow(\bs r),~c^\dagger_\downarrow(\bs r),~-c^\dagger_\uparrow(\bs r)\right]^T$ reads
\begin{align}
    H_{\rm tot}(t)=\frac{1}{2}\int d{\bs r}\,\Psi^\dagger({\bs r})H_{\rm BdG}(t)\Psi({\bs r})\,,
\end{align}
where, as described in the main text [Eq.~\eqref{eq:1}], the BdG Hamiltonian is
\begin{align}
\label{teq:1}
 H_{\rm BdG}(t)&=H_0+\sum_{j=1,2}V_{j}(t)\delta({\bs r-\bs R}_j)\,,\\
H_0&=\epsilon_p\tau_z+\Delta\tau_x\,,\nn\\
V_j(t)&=J_j{\bs S_j(t)}\cdot{\bs\sigma}\,,
\end{align}
while ${\bs S}_i(t)=S(\sin\theta_i\cos{\phi_i}, \sin\theta_i\sin{\phi_i}, \cos{\theta_i})$, 
with $\theta_i$ and $\phi_i$ being the (time-dependent) polar and azimuthal angles for spin $i=1,2$. 
  The time dependent Schrodinger equation can be written as $i\partial_t\psi(\bs r,t)=H_{\rm BdG}\psi(\bs r,t)$. By using  the Fourier decomposition $\psi({\bs r},t)= \frac{1}{L^d}\sum_{\bs k}e^{-i{\bs k}\cdot{\bs r}}\psi({\bs k},t)$, we can recast the BdG equation in the following  form:
\begin{align}
\label{teq:2}
\left(i\frac{\partial}{\partial t}-H_0({\bs k})\right)\psi({\bs k},t)=\sum_{j=1,2}V_j(t)\psi({\bs r}_j,t)e^{i{\bs k}\cdot{\bs r}_j}\,,
\end{align}
which, in the static limit pertains to the substitution $i\partial_t\rightarrow E$ and coincides with the equation for the spectrum presented in Ref.~\cite{pientka2013}. In the frequency domain, and retaining only the leading order terms in $\omega$, we obtain:
\begin{align}
\label{teq:3}
\psi({\bs r}_i,\omega)&
\approx-\sum_{j=1,2}[(\omega+\Delta\tau_x)I_0({\bs r}_j-{\bs r}_i)+I_1({\bs r}_j-{\bs r}_i)\tau_z]\int d\omega'V_j(\omega')\psi({\bs r}_j,\omega-\omega')e^{i{\bs k}\cdot{\bs r}_j}\,,
\end{align}
where
\begin{equation}
\label{teq:4}
I_0({\bs r})\equiv I_0(r) =\int\frac{d{\bs k}}{(2\pi)^d}\frac{e^{i{\bs k}\cdot{\bs r}}}{\epsilon_k^2+\Delta^2}\,;\,\,\,I_1({\bs r})\equiv I_1(r)=\int\frac{d{\bs k}}{(2\pi)^d}\frac{\epsilon_k e^{i{\bs k}\cdot{\bs r}}}{\epsilon_k^2+\Delta^2}\,.
\end{equation}
In the above expressions, we retained only the leading order corrections in $\omega$, assuming that the time dynamics of the classical spins as well as that of the emerging Shiba energies is such that $\omega\ll\Delta$ (adiabatic regime). For a 2D superconductor the integrals $I_{0,1}(r)$ can be written as $I_0(r)=(2\nu_0/\Delta)\tilde{I}_0(r)$ and $I_1(r)=2\nu_0\tilde{I}_1(r)$ \cite{pientka2013}, where   
 \begin{align}
\tilde{I}_{0(1)}(r)={\rm Im(Re)~K_0}\left[-i\left(1+i\frac{\Delta}{v_F k_F}\right)k_Fr\right]\,,
\end{align}
 $\nu_0$ is the density of states,  $k_F$ the Fermi momentum, $v_F$ the Fermi velocity  and  $K_0$ is the modified Bessel function of second kind. At $r=0$ they give $\tilde{I}_{0}=\pi/2$ and $\tilde{I}_{1}=0$, while  
for $k_Fr\gg1$ (a limit utilized throughout our work) the asymptotic expressions are \cite{kaladzhyan2017}
 \begin{align}
\tilde I_0(r)\approx\sqrt{\frac{2}{\pi}}\frac{\sin\left(k_Fr+\pi/4\right)}{\sqrt{k_Fr}}e^{-\Delta r/v_F}\,,~~
\tilde I_1(r)\approx\sqrt{\frac{2}{\pi}}\frac{\cos\left(k_Fr+\pi/4\right)}{\sqrt{k_Fr}}e^{-\Delta r/v_F}\,.
 \end{align}
Next, we switch back to the time-domain, and we get 
\begin{align}
\label{teq:5}
\psi({\bs r}_i,t)&\approx-\sum_{j=1,2}[(i\partial_t+\Delta\tau_x)\tilde{I}_0({\bs r}_j-{\bs r}_i)+\tilde{I}_1({\bs r}_j-{\bs r}_i)\tau_z]V_j(t)\psi({\bs r}_j,t)\,.
\end{align}
Defining $\tilde{\psi}({\bs r}_i,t)=\alpha_i\psi({\bs r}_i,t)$, we can manipulate further this expression by writing the combined evolution as:
\begin{align}
\label{teq:6}
    i\frac{\partial}{\partial t}\left(
    \begin{array}{c}
    \tilde{\psi}({\bs r}_1,t)\\
    \tilde{\psi}({\bs r}_2,t)
    \end{array}
\right)=\left(
\begin{array}{cc}
H_{1}(t) & H_{T}(t)\\
H^\dagger_{T}(t) & H_{2}(t)
\end{array}
\right)\left(
\begin{array}{c}
    \tilde{\psi}({\bs r}_1,t)\\
    \tilde{\psi}({\bs r}_2,t)
    \end{array}
\right)\,,
\end{align}
where
\begin{align}
    H_{i}(t)&=-\Delta\left(\frac{{\bs n}_i\cdot{\bs \sigma}}{\alpha_i}+\tau_x\right)+({\bs n}_i\times\dot{{\bs n}}_i)\cdot{\bs \sigma}\,,\nn\\
    H_{T}(t)&=-\Delta({\bs n}_1\cdot{\bs \sigma})({\bs n}_2\cdot{\bs \sigma})[\tilde I_0(R) \tau_x+\tilde I_1(R)\tau_z]\,,
    \label{teq:7}
\end{align}
and ${\bs n}_i\equiv {\bs n}_i(t)={\bs S}_i(t)/S_i$. As mentioned in the main text, the dynamics of the spins induce an extra term in the local Hamiltonian matrix element of Berry phase origin. Without this term, the transverse spin expectation values at the positions of the impurities would have the wrong sign.

To help distinguish the low and high energy sectors, which in turn will allow us to eliminate perturbatively the terms that couple them, it is instructive to perform first a unitary transformation $U_0=\exp{(i\pi\tau_y/4)}$ that converts $\tau_x\leftrightarrow\tau_z$, followed by a (time-dependent) $U_{i}(t)=\tau_0\otimes\tilde{U}_{i}(t)$ that acts on site $i=0(R)$ and diagonalizes the terms $\propto {\bs n}_i\cdot{\bs \sigma}$:  
\begin{align}
\tilde{U}_i(t)=\left(
\begin{array}{cc}
    \cos{\theta_i/2} & \sin{\theta_i/2}\\
    e^{i\phi_i}\sin{\theta_i/2} & -e^{i\phi_i}\cos{\theta_i/2}
\end{array}
\right)\,. 
\label{teq:9}
\end{align}
These rotations affect the terms in Eq.~(\ref{teq:7}) and they become:
 \begin{align}
\widetilde{H}_{i}(t)&=\tilde{U}_i^\dagger H_i(t)\tilde{U}_i-i\tilde{U}_i^\dagger\dot{\tilde{U}}_i=-\Delta\left(\frac{1}{\alpha_i}\sigma_z+\tau_z\right)+\frac{\dot{\phi}_i}{2}(1-\cos{\theta_i}\sigma_z+\sin\theta_i\sigma_x)-\frac{\dot{\theta}_i}{2}\sigma_y\,,\nn\\
\widetilde{H}_{T}(t)&=\tilde{U}_1^\dagger H_T(t)\tilde{U}_2=-\Delta\, U_1^\dagger({\bs n}_1\cdot{\bs \sigma})({\bs n}_2\cdot{\bs \sigma})U_2\,[\tilde I_0(R) \tau_z-\tilde I_1(R)\tau_x]\,.
\label{teq:8}
\end{align}
The low (high) $4\times4$ energy sector is spanned by the $\sigma_z\tau_z=-1(1)$ and the corresponding energies of the isolated Shiba states are  $\pm\Delta(1-1/\alpha_i)$ [$\pm\Delta(1+1/\alpha_i)$].  Consequently, we can then project the remaining terms, i.~e.,~the tunneling and the velocity contributions $\propto\dot{\phi}_i,\dot{\theta}_i$, onto the low-energy sector to obtain an effective time-dependent Hamiltonian. 

To simplify the discussion, from here onward, we assume the left  spin ($1$) is static and aligned along the $z$-direction, or $\theta_1,\phi_1=0$, and that  $\theta_2\equiv\theta$ and $\phi_2\equiv \phi$. Furthermore, we also consider that all lengths are expressed in terms of $\xi_{SC}=v_F/\Delta$, and set $\Delta=1$.  
Then, the projected $4\times 4$ low-energy Hamiltonian can be written as
\begin{align}
    H_l(t)\equiv \mathcal{P}_l\tilde{H}(t)\mathcal{P}_l=H_{l,0}(t)+\dot{\phi}A_{l,\phi}(t)+\dot{\theta}A_{l,\theta}(t)\,,
    \label{teq:9a}
    \end{align}
where $\mathcal{P}_l$ is the corresponding projector while
 {\small       \begin{equation}
    H_{l,0}(t)=\begin{bmatrix}
    -1+\frac{1}{\alpha_1} &0& -\frac{e^{-R+i\phi}\cos(\theta/2)\sin(k_FR+\pi/4)}{\sqrt{\pi k_FR/2}} &  \frac{e^{-R+i\phi}\sin(\theta/2)\cos(k_FR+\pi/4)}{\sqrt{\pi k_FR/2}}\\
    0  & 1-\frac{1}{\alpha_1} &-\frac{e^{-R}\sin(\theta/2)\cos(k_FR+\pi/4)}{\sqrt{\pi k_FR/2}} &  \frac{e^{-R}\cos(\theta/2)\sin(k_FR+\pi/4)}{\sqrt{\pi k_FR/2}}\\
    -\frac{e^{-R-i\phi}\cos(\theta/2)\sin(k_FR+\pi/4)}{\sqrt{\pi k_FR/2}} &  -\frac{e^{-R}\sin(\theta/2)\cos(k_FR+\pi/4)}{\sqrt{\pi k_FR/2}} & -1+\frac{1}{\alpha_2} &0\\
        \frac{e^{-R-i\phi}\sin(\theta/2)\cos(k_FR+\pi/4)}{\sqrt{\pi k_FR/2}} &  \frac{e^{-R}\cos(\theta/2)\sin(k_FR+\pi/4)}{\sqrt{\pi k_FR/2}} &0 & 1-\frac{1}{\alpha_2}\,,
    \end{bmatrix}
    \label{teq:10}
    \end{equation}}
represents the instantaneous projected Hamiltonian, with  $A_{l,\phi}(t)$ and $A_{l,\theta}(t)$ being the projected gauge field terms associated with the $\dot{\phi}$ and $\dot{\theta}$ contributions in Eq.~(\ref{teq:9a}).

\section{\label{app:2}Effective qubit Hamiltonian}

We can further diagonalize the instantaneous $4\times4$ Hamiltonian $H_{l,0}(t)$ in order to identify the effective $2\times2$ YSR qubit Hamiltonian presented in the main text. That is achieved by another time-dependent unitary transformation
\begin{equation}
    U_3(t)=\frac{1}{2\sqrt{BC}}\begin{bmatrix}
     -s_2 B_{+}C_{-}e^{i\phi} & s_2 B_{-}C_{+}e^{i\phi} & -s_2B_+C_+e^{i\phi} & s_2B_{-}C_{-}e^{i\phi}\\
      s_1B_{-}C_{+} &  - s_1B_{+}C_{-} & -s_1B_{-}C_{-} & s_1 B_+C_+\\
     s_1s_2C_{+}B_{+} & s_1s_2C_{-}B_{-} & -s_1s_2C_{-}B_+ & -s_1s_2B_{-}C_+\\
     B_{-}C_{-} &  B_{+}C_{+} &  B_{-}C_+ & C_{-}B_+
    \end{bmatrix}
     \label{teq:12}
\end{equation}
where
\begin{align}
t_h=&\frac{4\alpha_1\alpha_2}{\sqrt{2\pi k_FR}}e^{-R}\,,\nn\\
 B=&\sqrt{\left(2\alpha_2\alpha_1-\alpha_1-\alpha_2\right)^2+\left(t_h\cos(k_FR+\pi/4)\sin(\theta/2)\right)^2} \,,\nn\\
 C=&\sqrt{(\alpha_1-\alpha_2)^2+\left(t_h\sin(k_FR+\pi/4)\cos(\theta/2)\right)^2}\,,\nn\\
 B_\pm=&\sqrt{B\pm\left(2\alpha_1\alpha_2-\alpha_2-\alpha_1\right)}\,,\nn\\
 C_\pm=&\sqrt{C\pm(\alpha_1-\alpha_2)}\,,
  \label{teq:13}
\end{align}
with $s_1=\mbox{sign}[\sin(k_FR+\pi/4)]$ and $s_2=\mbox{sign}[\cos(k_FR+\pi/4)]$. Its effect on $H_{l}(t)$ can be formally written as:
\begin{align}
    \tilde{H}_{l}(t)&=U_3^\dagger H_{l}(t)U_3-iU_3^\dagger\dot{U}_3=\tilde{H}_{l,0}(t)+\dot{\phi}\tilde{A}_{l,\phi}(t)+\dot{\theta}\tilde{A}_{l,\theta}(t)\,,\\
    \tilde{H}_{l,0}(t)&=U_3^\dagger H_{l,0}(t)U_3\,;\,\,\,\tilde{A}_{l,s}(t)=U_3^\dagger A_{l,s}U_3-iU_3^\dagger\partial_s U_3\,,
\end{align}
with $s=\phi,\theta$. Note that while $\tilde{H}_{l,0}(t)$ is now diagonal, with energies $\pm E_{1,2}$, where $E_{1,2}=(B\pm C)/(2\alpha_1\alpha_2)$, the gauge field terms can  induce transitions between its eigenstates. More importantly, these terms have two separate contributions: one from the initial gauge fields, originating from the first unitary transformations $U_{1,2}$, and one from the diagonalization of the $4\times4$ effective (instantaneous) time-dependent Hamiltonian $H_{l,0}(t)$. Both are required to correctly capture the low-energy sector dynamics, and starting from the static effective theory by turning the parameters $\theta$ and $\phi$ time-dependent would lead to erroneous results. 

In order to establish a qubit that is well separated from the excited states,  we assume that both $\alpha_1,\alpha_2\neq 1$, as well as $t_h,\delta\alpha\ll2{\rm min}|1-\alpha_{1,2}|$, with $\delta\alpha=|\alpha_1-\alpha_2|$. Then,  we can further project $\tilde{H}_l(t)$ to the two lowest energy states, resulting in the qubit Hamiltonian presented in the main text (up to terms that act as identity in this subspace)
\begin{align}
  \label{seq:12}
    H_q(t)=&\frac{\epsilon_q}{2}\Sigma_z+\bs\beta(t)\cdot\bs\Sigma\,,\\
    \beta_{z}(t)=&\frac{\alpha_2-\alpha_1}{2\alpha_1\alpha_2\epsilon_q}\sin^2(\theta/2)\,\dot\phi\,,\nn\\
    \beta_{x}(t)=& t_h\frac{\sin(k_FR+\pi/4)\sin\theta\sin(\theta/2)}{4\alpha_1\alpha_2\epsilon_q}\,\dot\phi\,,\nn\\
    \beta_{y}(t)=& t_h\frac{(\alpha_1-\alpha_2)\sin(k_FR+\pi/4)\sin(\theta/2)}{4(\alpha_1\alpha_2)^2\epsilon_q^2}\,\dot\theta\,,
\label{teq:14}
\end{align}
where $\epsilon_q=C/\alpha_1\alpha_2$ is the qubit splitting energy.

\section{\label{app:3}Details on the read out via torques}

Here we provide details on the calculation of the torque  ${\bs \tau}=-J_1S{\bs n}_1\times\langle{\bs \sigma}(0)\rangle$ acting on the precessing spin ${\bs S}_1$ by the SC electrons, and its effects on the STM-ESR signal. We first note that at the operator level, the torque can be written as $\hat{\bs \tau}=-{\bs n}_1\times\hat{\bs h}$, where we introduced the magnetic field operator $\hat{\bs h}(t)=\partial_{{\bs n_1}}H_{\rm tot}(t)$. Then, for a given many-body state $|\Psi(t)\rangle$ that acts in the occupation number basis we have
\begin{align}
    {\bs h}_\Psi=\langle\Psi(t)|\hat{\bs h}|\Psi(t)\rangle=\langle\Psi(t)|\partial_{{\bs n_1}}H_{\rm tot}(t)|\Psi(t)\rangle\,.
\end{align}
For a static BdG Hamiltonian, we can write $H_{\rm tot}=\sum_{i}\epsilon_i({\bs n}_i)(\gamma_i^\dagger\gamma_i-1/2)$, with $\epsilon_i({\bs n}_1)$ being the BdG eigenvalues, and $\gamma_i$ ($\gamma_i^\dagger$) being the Bogoliubov annihilation (creation) operator found from diagonalization. Then, in such a case we obtain the average field:
\begin{align}
    {\bs h}=\langle\hat{\bs h}\rangle=\sum_{i}(f_i-1/2)\partial_{{\bs n}_1}\epsilon_i({\bs n}_1)\,,
\end{align}
with $f_i=\langle \gamma_i^\dagger\gamma_i\rangle$ being the occupation of state $i$. This   encodes both the well-known RKKY interaction mediated by the bulk states, as well as the (static) YSR contribution \cite{YaoPRL14}. Dynamics can induce transitions between different instantaneous energy levels, and in general a full diagonal form for the BdG Hamiltonian might not be found. However, in our perturbative scheme in the dynamics, when $\epsilon_q\neq2\beta_z$, we can neglect the transitions caused by $\beta_x$ and $\beta_y$. Moreover, since $\Omega_1\ll\Delta$, the bulk states are also unaffected. Then, the many-body Hamiltonian is still diagonal, and the  magnetic field  reads:
\begin{equation}
    {\bs h}=-\frac{1}{2}\sum_{i\in {\rm bulk}}\partial_{{\bs n}_1}\epsilon_i({\bs n}_1)+\sum_{i=1,2}(f_i-1/2)\partial_{{\bs n}_1}\epsilon_i^{\rm eff}({\bs n}_1)\equiv {\bs h}_{\rm bulk}+{\bs h}_{\rm YSR}\,,
\end{equation}
where the first and second terms determine the bulk contribution (all levels $i$ empty, or $f_i=0$) and YSR in-gap states contributions, respectively. Importantly, $\epsilon_i^{\rm eff}$ are the full single-particle energies that include the shifts induced by the dynamics (which, in a more formal language, corresponds to Berry phase effects \cite{mishra2020}). The YSR states that define the qubit states correspond in the many-body picture to the configurations $f_{1(2)}=0(1)$ and $f_{1(2)}=1(0)$. Thus, the field for each qubit state is
\begin{align}
    {\bs h}_\sigma={\bs h}_{\rm bulk}+\frac{(-1)^\sigma}{2}\partial_{{\bs n}_1}[\epsilon_1^{\rm eff}({\bs n}_1)-\epsilon_2^{\rm eff}({\bs n}_1)]\equiv{\bs h}_{\rm bulk}+\frac{(-1)^\sigma}{2}({\bs h}_{s}+{\bs h}_d)\,,
\end{align}
with $\sigma=0,1$. Consequently, one can find  the corresponding torques from ${\bs \tau}_\sigma=-{\bs n}_1\times{\bs h}_\sigma$,  ${\bs \tau}_s=-{\bs n}_1\times\partial_{{\bs n}_1}\epsilon_q$ and ${\bs \tau}_d=-2{\bs n}_1\times\partial_{{\bs n}_1}\beta_z$, as  presented in the main text (note that these are more general as they assume arbitrary changes in the angles $\theta$ and $\phi$).

For the numerical evaluation of the torques we have employed a Green function approach that describes the dimer when the target spin precesses circularly. In this case, an exact solution can be found, assuming  $\phi(t)=\Omega t$ and $\theta=$const, with $\Omega$ being the precession frequency. Indeed,  the dynamical problem in Eq.~\eqref{teq:1} can be made static by rotating it with the time-dependent unitary transformation $\mathcal{U}(t)=e^{-i(\Omega/2)\sigma_zt}$. In this frame, the stationary Schrodinger equation from Eq.~\eqref{teq:1} can be written as 
 \begin{equation}
\tilde H_{\rm BdG}=H_{\rm BdG}(0)-\frac{\Omega}{2}\sigma_z\,,
\label{teq:15}
 \end{equation}
 where the second term acts as a fictitious magnetic field on the superconductor. Following Ref.~\onlinecite{kaladzhyan2017}, the wave-function at any point $\bs r$ can be written as, $\psi({\bs r})= \sum_{r_j\in 0,R} G_0 ({\bs r}-{\bs r}_j,E)V_j\psi({\bs r_j})$, where 
  \begin{equation}
G_0(\bs r,E)=-\begin{bmatrix}
(E+\Omega/2) I^+_0+I^+_1 & 0 & I^+_0 & 0\\
0 &(E-\Omega/2) I^-_0+I^-_1 & 0 & I^-_0 \\
I^+_0 & 0 & (E+\Omega/2) I^+_0-I^+_1 & 0\\
0 & I^-_0 & 0 & (E-\Omega/2) I^-_0-I^-_1
\end{bmatrix}
\,,
\label{teq:16}
 \end{equation}
 with 
 \begin{align}
 I^\pm_0(r)=\int \frac{d{\bs k}}{(2\pi)^d} \frac{e^{i\bs k\cdot\bs r}}{1+\epsilon^2_k-(E\pm \Omega/2)^2}\,,~~
I^\pm_1(r)=\int \frac{d{\bs k}}{(2\pi)^d} \frac{\epsilon_k e^{i\bs k\cdot\bs r}}{1+\epsilon^2_k-(E\pm \Omega/2)^2}\,.
\label{teq:17}
 \end{align}
We then  get the following set of eigenvalue equations: 
\begin{align}
   \left(1-G_0({ 0},E)V_1\right)\psi({0})&=G_0({ R},E)V_2\psi({ R})\,,\nn\\
  \left(1-G_0 ({ 0},E)V_2\right)\psi({R})&=G_0({ R},E)V_1\psi({ 0})\,,
  \label{teq:18}
\end{align}
and  the in-gap spectrum can be found numerically for arbitrary frequencies $\Omega/2<1$ from the $8\times8$ determinant
\begin{align}
    \left|\begin{array}{cc}
    1-G_0({ 0},E)V_1 & G_0({ R},E)V_2\\
    G_0({ R},E)V_1 & 1-G_0 ({0},E)V_2
    \end{array}
    \right|=0\,.
\end{align}
Then, the associated torques (that include all orders in $\Omega/2$) can be evaluated as in the previous subsection. The plots depicted in Fig.~\ref{fig:3} in the main text were obtained assuming the deep Shiba limit ($\alpha_{1,2}\sim1$) and  $\Omega/2\ll1$ (adiabatic driving), which is the relevant regime in this work. Nevertheless, this approach can be readily employed to study the effects of the dynamics beyond the adiabatic realm.

\section{\label{app:4}Linearization of LLG equation and resonance frequency renormalization}

The LLG equation describing the dynamics of the classical spin $\bs S_1$ in the presence of the torque ${\bs \tau}_\sigma$ torque pertaining to the YSR qubit in state $\sigma=0,1$ can be written as 
\begin{equation}
    \dot{\bs S_1}(t)=-\gamma\bs S_1(t)\times\bs B(t)+\bs\tau_\sigma(t)+\alpha_g \bs S_1\times \dot{\bs S_1}\,,
\end{equation}
where $\bs B(t)=B_0\bs z+\bs B_\perp(t)$ is the external magnetic field, being the sum of a constant term along $z$ which defines the bare resonance frequency $\Omega_0=\gamma B_0$, and a weak in-plane rf component.
Specifically, we consider $\bs B_\perp(t)=B_\perp(\cos(\Omega_1 t),\sin(\Omega_1 t),0)$ and $B_\perp\ll B_0$. In the stationary limit,  the impurity spin can be written as ${\bs S_1}(t)=S_z {\bs z}+\delta{\bs S}(t)$, with  $\delta{\bs S}(t)=S_\perp\left(\cos(\Omega_1 t+\phi\right),\sin\left(\Omega_1 t+\phi), 0\right)$, where $S_\perp=S\sin\theta\approx S\theta$ and $S_z=S\cos{\theta}\approx S$, and $\phi$ quantifies the lagging of the spin with respect to the driving field. In this limit, we can also expand the torque ${\bs \tau}_\sigma$ in terms of the small parameter $\theta$, which in turn gives \cite{mishra2020}: 
\begin{align}
    \left[\alpha_g\Omega_1 S+\left(\Omega_1-\Omega_0+\frac{\tau'_{\sigma s}}{S}+\Omega_1\frac{\tau'_{\sigma d}}{S}\right){\bs z}\times\right]\delta{\bs S}(t)&\approx-\gamma S {\bs z}\times {\bs B}_\perp(t)\,,
    \label{LLG}
\end{align}
where 
$\tau'_{\sigma s}=\left(\partial\tau_{\sigma s}/\partial\theta\right)|_{\theta=0}$ and $\tau'_{\sigma d}=\left(\partial\tau_{\sigma b}/\partial\theta\right)|_{\theta=0}$ with $\tau_{\sigma b}=\left(\partial\tau_{\sigma d}/\partial\Omega_1\right)|_{\Omega_1=0}$. 
     
From the above equations, we can readily evaluate both the amplitude $S_\perp$ and the phase lag $\phi$, respectively: 
\begin{align}
S_\perp&=\frac{\gamma S B_\perp}{\sqrt{(\Omega_1-\Omega_0+\tau'_{\sigma s}/S+\tau'_{\sigma d}\Omega_1/S)^2+(\alpha_gS\Omega_1)^2}}\,,\\
\phi&=\arctan\frac{\alpha_gS\Omega_1}{\Omega_1-\Omega_0+\tau'_{\sigma s}/S+\tau'_{\sigma d}\Omega_1/S}\,.
\end{align}
The resonance frequency $\Omega_{r,\sigma}$ of the precessing spin is shifted depending on the qubit state $\sigma=0,1$ as 
\begin{align}
    \Omega_{r,\sigma}=\frac{\displaystyle{\Omega_0-\frac{\tau'_{\sigma s}}{S}}}{\displaystyle{1+\frac{\tau'_{\sigma d}}{S}}}\approx\Omega_0\left(1-\frac{\tau'_{\sigma d}}{S}\right)-\frac{\tau'_{\sigma s}}{S}\,.
\end{align}
Note that each type of torque will also contain a constant contribution, independent of the qubit state, that originates from the (occupied) bulk states. Hence, we can write $\tau'_{\sigma s}\rightarrow\tau'_{bs}+\tau'_{\sigma s}$ and $\tau'_{\sigma d}\rightarrow\tau'_{bd}+\tau'_{\sigma d}$, where the index $b$ labels bulk contribution. Nevertheless, as showed in Ref.~\cite{YaoPRL14} the static bulk contribution is negligible for $R\geq1$, while as argued in Ref.~\cite{mishra2020}, the dynamical contribution of the bulk states is negligible in the adiabatic regime.  We can then extract the resonance frequency difference as
\begin{align}
    \delta\Omega=\Omega_{r,0}-\Omega_{r,1}\approx\frac{1}{S}\left(\Omega_0\tau'_{d}+\tau'_{s}\right)\,,
\end{align}
which reflects only the in-gap state effects.

\section{\label{app:5}Manipulation of the YSR qubit: Rabi oscillations}

\subsubsection{Behavior of Rabi 
oscillation period around $\theta=0$ and $\pi$}
      
\begin{figure}
\centering          \includegraphics[width=0.9\linewidth]{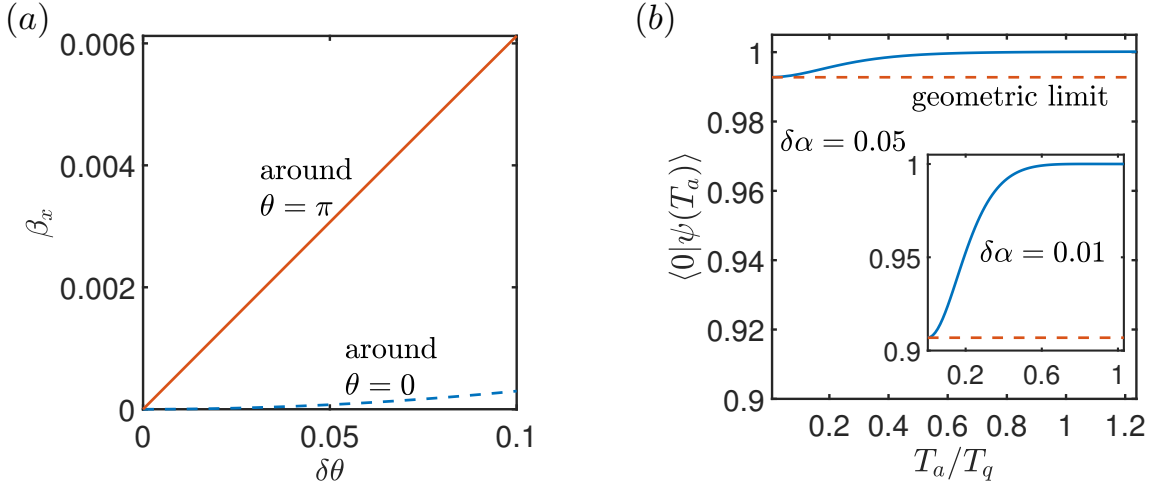}
\caption{($a$) The Rabi oscillation strength $\beta_x$ in  Eq.~\eqref{eq:7} as a function of the  deviation $\delta\theta$ around $\theta=0$ (blue dashed line) and $\pi$ (red solid line) for $\alpha_1=1.15$, $\alpha_2=1.1$, and $R=2.9$. Note that for these parameters, $t_h/\delta\alpha=0.34$. ($b$) The probability amplitude of the qubit initialized in state $|0\rangle$ to remain in that state after time $T_a$, $\langle 0|\psi(T_a)\rangle$, in terms of $T_q=\hbar/\epsilon_q$ for $\delta\alpha=0.05$, or $t_h/\delta\alpha=0.34$ (inset $\delta\alpha=0.01$, or $t_h/\delta\alpha=1.75$). The blue curve is the result obtained from the full numerical implementation of the evolution, while the red-dashed line is $\langle 0|\psi(T_a)\rangle$ evaluated in the geometrical limit using Eq.~\eqref{seq:1}. As $T_a$ is increased, the full curve starts deviating from the geometric limit valid for $T_a\ll T_q$, and reaches almost unity already for $T_a\sim T_q$, consistent with the adiabatic result. All plots are obtained using $k_F=13.55$.}
\label{fig:S1}
\end{figure}

The Rabi frequency and hence, the time period of the Rabi oscillation is determined by $\beta_x$. Notably, the $\beta_i$ terms are much weaker for deviation $\delta\theta$ around $\theta=0$ as compared to such deviation around $\theta=\pi$ making it inefficient for manipulation in the parallel configuration. In   Fig.~\ref{fig:S1}a we show $\beta_x$ as a function of $\delta\theta$ near  $\theta=0$ and near $\pi$,  suggesting that $\beta_x$ varies linearly (quadratically) around $\theta=\pi$ ($\theta=0$).
      
\subsection{Numerical approach for qubit state evolution}
      
The time evolution operator corresponding to the qubit Hamiltonian can be written as $U(t,t_0)=\mathcal{T}e^{-\frac{i}{\hbar}\int^{t}_{t_0}dt'H_q(t')}$, where $\mathcal{T}$ represents the time ordering operator. We have implemented the  evolution of the qubit state by performing time slicing with small increment $\delta t$, so that the evolution operator during one slice can be expanded as
$U(t,t-\delta t)=1-(i/\hbar)H_q(t)\delta t$.
Then, starting from the initial state $|0\rangle$, the qubit state at time $t$ can then be written as
\begin{equation}
|\psi(t)\rangle=U(t,0)|0\rangle=U(t, t-\delta t)U(t-\delta t,t-2\delta t)\cdots U(\delta t,0)|0\rangle\,,    
\end{equation}
which we evaluate numerically for $|\psi(t)\rangle$ by evolving the state under the sequence of pulses described in the main text.
   
\subsection{Analytical approach for qubit state evolution in the geometric regime}

The qubit state evolution subjected to  the pulse $\theta(t)=\pi\tanh(2\pi t/T_a)$ can be studied analytically in two extreme limits: the adiabatic ($T_a\gg T_q$) and geometric ($T_a\ll T_q$) limit, respectively. In 
           the adiabatic limit, the qubit evolution is trivial, as it remains in state $|0\rangle$ during the pulse. In the geometric limit, the energy splitting $\epsilon_q$  becomes unimportant (thus can be neglected), and the qubit evolution is solely determined by $\beta_y\Sigma_y\propto\dot{\theta}$. Then, the  evolution operator under an arbitrary rotation of the qubit from $\theta=0$ to a final  $\theta_0$ reads $U_g(\theta_0)=e^{-i\int \beta_y\Sigma_y dt}\equiv \cos{A(\theta_0)}-i\sin{A(\theta_0)}\,\Sigma_y$, where 
               \begin{align}
    A(\theta_0)=&\int^{\theta_0}_0 d\theta\frac{\delta\alpha t_h\sin(k_FR+\pi/4)\sin(\theta/2)}{4\left(\delta\alpha^2+(t_h\sin(k_FR+\pi/4)\cos(\theta/2))^2\right)}\nn\\
    =&\frac{1}{2}\left[\tan^{-1}\left(\frac{t_h\sin(k_FR+\pi/4)}{\delta\alpha}\right)-\tan^{-1}\left(\frac{t_h\sin(k_FR+\pi/4)\cos(\theta_0/2)}{\delta\alpha}\right)\right]\,,
    \label{seq:1}
\end{align}
and thus the first pulse in Fig.~\ref{fig:4}a corresponds to $\theta_0=\pi$ in the geometric limits.             
           
           Fig:~\ref{fig:S1}b shows the probability amplitude of the qubit state to be in $|0\rangle$ as a function of time $T_a$  (scaled with $T_q$), $c_0(T_a)$, starting from $|0\rangle$. The blue solid line represents $c_0$ evaluated numerically while the red dashed line corresponds to the geometric limit evaluated as $c_0=\cos A(\pi)$. The deviation of the geometric amplitude  from adiabatic result increases with decreasing  $\delta\alpha$, which is due to the effect of tunneling that makes it easier for the qubit to explore the Bloch sphere. For $\delta\alpha>\epsilon_q$, the deviation is negligible and $c_0 (T_a)$ depends weakly on the pulse length $T_a$. Thus, this situation is preferable  for the qubit manipulation.

\subsection{Geometric effects around $\theta=\pi$}

\begin{figure}
    \centering
    \includegraphics[width=0.4\linewidth,trim=110mm 00mm 1mm 00mm,clip]{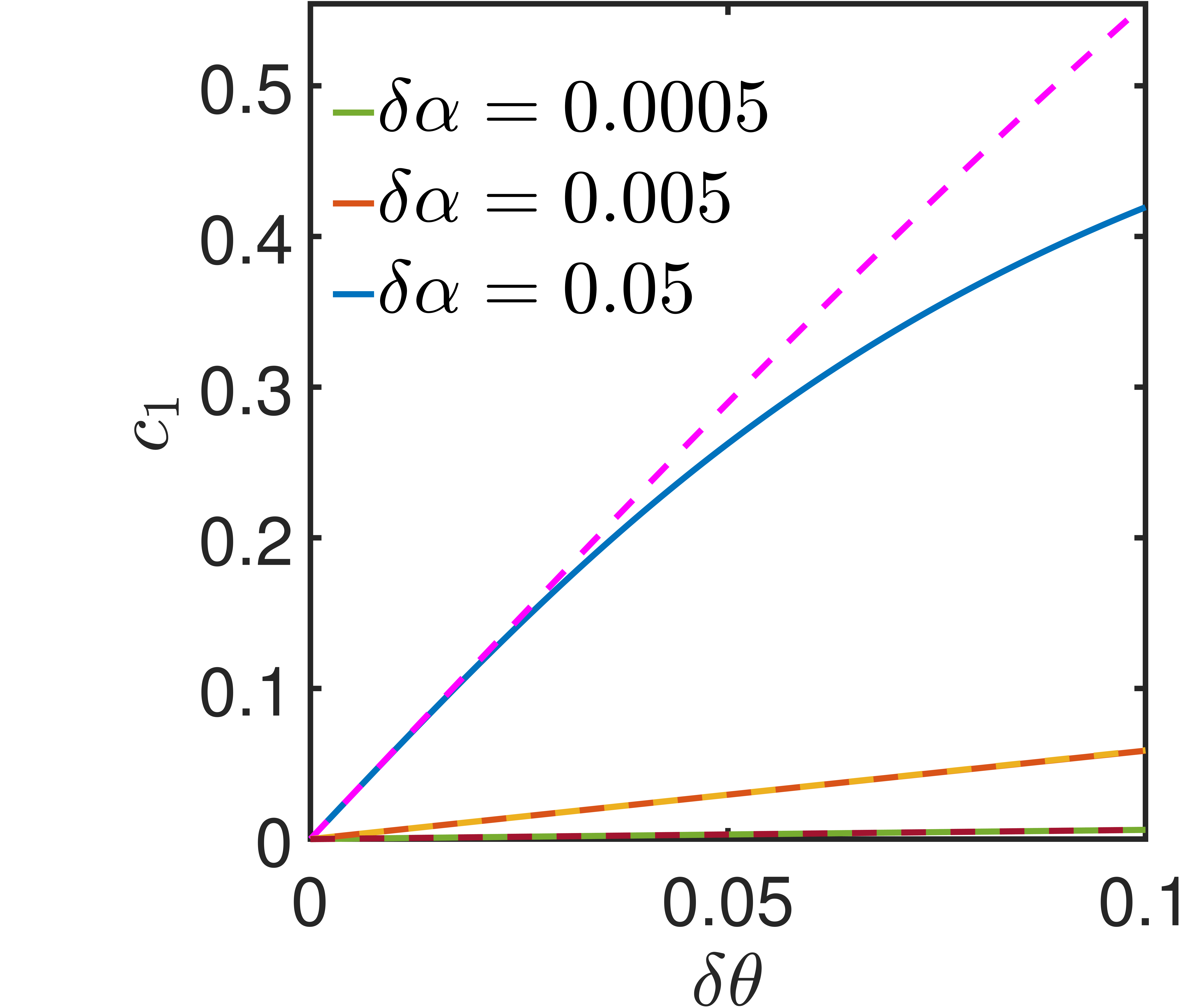}
        \caption{The dependence of  $c_1$ on $\delta\theta$ near $\theta=\pi$ for $\delta\alpha=0.0005,0.005,0.05$, and which corresponds to  $t_h/\delta\alpha=34,3.4,0.34$ at $R=2.9$. We see that for $\delta\alpha$ such that $t_h/\delta\alpha\leq 1$, the weight  $c_1$ increases linearly with $\delta\theta$, matching well the geometric limit in Eq.~\eqref{geq:1}, represented by the dashed lines, for a wide range of deviations $\delta\theta$.}
    \label{fig:S2}
  \end{figure}

The Rabi oscillations amplitude is also reduced because the geometric pulse $\theta(t)=(\pi-\delta\theta)\tanh(2\pi t/T_b)$ leads to a probability amplitude $c_1$ to excite the qubit from state $|0\rangle$ to state $|1\rangle$. Assuming that $t_h<\delta\alpha$ and  $\delta\theta \ll 1$, we obtain
\begin{equation}
    c_1\approx\frac{t_h\sin\left(k_FR+\pi/4\right)}{4\delta\alpha}\delta\theta\,.
    \label{geq:1}
\end{equation}
This linear increase of $c_1$ with $\delta\theta$ shows good agreement with the full numerical results, as shown in Fig.~\ref{fig:S2} for various values of $\delta\alpha$.  In the limit $t_h\ll\delta\alpha$, $c_1\rightarrow0$ and the amplitude of Rabi oscillations approaches unity, as argued in the main text.

  %%%%%%%%%%%%%%%%%%%%%%%%%%%%%%%%%%%%%%%%%%%%%%%%%%%%%%%%%%
  %%%%%%%%%%%%%%%%%%%%%%%%%%%%%%%%%%%%%%%%%%%%%%%%%%%%%%%%%%
\section{{\label{app:6}Decoherence of YSR qubit}}
Below we give the detailed analysis of the decoherence in the YSR qubit induced due to the fluctuations in the magnetic moments and phonon coupling to the Shiba electrons.

\subsection{Magnon-induced decoherence}       

Here we provide details on the decoherence of the YSR qubit by the stochastic fluctuations in the magnetic moments orientations. The fluctuations of the spin $k=1,2$ can be accounted for by performing the substitution $\bs n_k\rightarrow\bs n_k+\delta \bs n_k(t)$,  where  $\delta \bs n_k(t)=\bs e_{k,1} \delta n_{k,1}+\bs e_{k,2}\delta n_{k,2}\perp{\bs n}_k$ describe the induced fluctuations of the magnetic moment perpendicular to the deterministic orientations $\bs n_k$. Here,  $\bs e_{k,i}$ and $\delta n_{k,i}$ with $i=1,2$ label the orthogonal fluctuations directions and the corresponding magnitudes, respectively,  with $\bs e_{k,1}=\frac{\bs z\times\bs n_k}{|\bs z\times\bs n_k|}$ and $\bs  e_{k,2}=\frac{\bs n_k\times (\bs z\times \bs n_k)}{|\bs n_k\times  (\bs z\times \bs n_k)|}$.

Then, the coupling between the qubit and the classical spins changes accordingly, $V\rightarrow V+\delta V(t)$, with
\begin{align}
    \delta V(t)=&\sum_{k=1,2}J_kS\,
    \delta{\bs n}_k(t)\cdot{\bs \sigma}\delta({\bs r}-{\bs r}_k)=\sum_{k=1,2}\frac{\partial H_{{\rm BdG}}}{\partial{{\bs n}_k}}\cdot\delta{\bs n}_{k}(t)\equiv\sum_{k=1,2}\hat{\bs h}_k\cdot\delta {\bs n}_{k}(t)\,,
\end{align}
where $\hat{\bs h}_k$ is the magnetic field operator acting on the electrons.   

Projecting the above Hamiltonian onto the qubit subspace, leads to the following extra contribution: 
\begin{align}
\delta H_{q}(t)&=\sum_{k}\sum_{\mu=1,2;\nu=x,y,z}\delta n_{k,\mu}(t)\chi_{k}^{\mu\nu}\Sigma_\nu\,,
\label{q_magnons}
\end{align}
where $\chi_{k}^{\mu\nu}$ represent the components of the  tensor coupling between the fluctuations and the qubit, and which can be extracted from  above:
\begin{align}
    \chi_{k}^{\mu\nu}=\frac{1}{2}{\rm Tr}[{\bs e}_{k,\mu}\cdot\hat{\bs h}_k\,\Sigma_\nu]\,,
\end{align}
with the trace being taken over the qubit states. In the following, we assume the external driving is absent, and only focus on the static coherence properties. Then, we can evaluate explicitly the matrix elements of the field:
\begin{align}
    \langle\sigma|\hat{\bs h}_k|\sigma\rangle&=-(-1)^{\sigma}\frac{1}{2}\partial_{\bs n_k}\epsilon_q=(-1)^{\sigma+k}\frac{1}{2}\partial_{\theta}\epsilon_q{\bs e}_{k,2}\,,\\
     \langle\sigma|\hat{\bs  h}_k|\bar{\sigma}\rangle&=(-1)^{\sigma}\epsilon_q\langle\sigma|\partial_{\bs n_k}\bar{\sigma}\rangle=-(-1)^{\sigma+k}\epsilon_q\left[\langle\sigma|\partial_{\theta}\bar{\sigma}\rangle{\bs e}_{k,2}+\frac{1}{\sin{\theta}}\langle\sigma|\partial_{\phi}\bar{\sigma}\rangle{\bs e}_{k,1}\right]\,,
\end{align}
where the factor $(-1)^k$ reflects that the magnetic fields are opposite for given relative angles. Note that for $\delta\alpha\neq0$, the diagonal terms vanish at both $\theta=0$ and $\theta=\pi$. From the above expressions, we can write the total magnetic field operator acting in the qubit subspace as:
\begin{align}
    \hat{\bs h}^q_k&=(-1)^{k}\left[\frac{1}{2}\partial_{\theta}\epsilon_q\Sigma_z{\bs e}_{k,2}-\epsilon_q\left(\beta_y'\Sigma_x{\bs e}_{k,2}+\frac{\beta_x'}{\sin{\theta}}\Sigma_y{\bs e}_{k,1}\right)\right]\,,
\end{align}
where $\beta'_{x}=\beta_x/\dot{\phi}$ and $\beta'_{y}=\beta_y/\dot{\theta}$ in the qubit Hamiltonian in Eq.~(\ref{teq:14}). From here, the matrix $\chi_k^{\mu\nu}$ can be readily identified. Let us evaluate the above field for the two cases of interest $\theta=0$ (parallel) and $\theta=\pi$ (anti-parallel) configurations, respectively. In the former case, $\hat{\bs h}^q_k\equiv0$, meaning that no dephasing or relaxation occurs because of the coupling to the magnetic fluctuations, while in the latter $\hat{\bs  h}^q_k=(-1)^{k+1}g_c\left(\Sigma_x{\bs e}_{k,2}+\Sigma_y{\bs e}_{k,1}\right)$ where 
\begin{align}
    g_c=\frac{t_h\sin(k_FR+\pi/4)}{4\alpha_1\alpha_2}\,,
\end{align}
is the effective coupling strength of the qubit to the fluctuations whose magnitude is dictated by the tunneling $t_h$. For this specific orientation, Eq.~(\ref{q_magnons}) becomes:
\begin{align}
    \delta H_q(t)&=g_c\sum_k(-1)^{k+1}\left[\delta n_{k,2}(t)\Sigma_x+\delta n_{k,1}(t)\Sigma_y\right]\,.
\end{align}
 We are now in position to calculate the decoherence rates engendered by this coupling.  We first introduce the noise power spectrum pertaining to the fluctuations $\delta{\bs n}_k(t)$ in the Fourier space:
 \begin{align}
     S^{kk'}_{\mu\nu}(\omega)=\frac{1}{2\pi}\int dt e^{-i\omega t}\langle \delta n_{k,\mu}(t)\delta n_{k',\nu}(0)\rangle\delta_{kk'}\equiv S^{k}_{\mu\nu}(\omega) \,,
 \end{align}
 where the averages are taken over the thermal equilibrium, and we assumed the fluctuations of the two spins are not correlated. Within the Bloch-Redfield framework \cite{blum2012}, the dephasing and the longitudinal relaxation rates read, respectively: 
\begin{align}
    \Gamma_{\phi,m}&=\sum_{k=1,2}(|\chi^{1z}_{k}|^2+|\chi^{2z}_{k}|^2)S^k_{11}(0)\,,\\
    \Gamma_{1,m}&=
    \sum_{\mu,\nu}\sum_{k=1,2; \sigma=\pm}\chi^{\mu \sigma}_{k}\chi^{\nu \bar{\sigma}}_{k}S^k_{\mu\nu }(\sigma\epsilon_q)\,,
\end{align}
where $\chi^{\mu \pm}_{k}=\chi^{\mu x}_{k}\pm  i\chi^{\mu y}_{k}$.  The pure dephasing rate $\Gamma_\phi$ vanishes at both $\theta=0$ and $\pi$, and $\Gamma_{1,m}=0$ at $\theta=0$. The relaxation rate at $\theta=\pi$ is      
\begin{align} 
\Gamma_{1,m}=1/T_{1,m}=2g_c^2\sum_{k=1,2}\left[S^k_{11}(\epsilon_q)+S^k_{11}(-\epsilon_q)-i(S^k_{12}(-\epsilon_q)-S^k_{12}(\epsilon_q))\right]\,,
\end{align}
while the dephasing time satisfies $T_{2,m}=2T_{1,m}$. In order to give estimates, we need to describe the noise spectrum of the magnetic fluctuations. To do that, we start by employing  the stochastic
LLG equation describing the magnets in the presence of magnetic noises (here we disregard the effect of the qubit on the dynamics, as it would only manifest in higher orders in the coupling):
\begin{align}
    \dot{\bs n}_k=-\gamma {\bs n}_k\times\bigg({\bs B}_{k,\rm eff}+\delta\bs{ B}_k(t)\bigg)+\alpha_g{\bs n}_k\times\dot{\bs n}_k\,.
\end{align}
Here  ${\bs B}_{k,\rm eff}=-\gamma^{-1}\delta F_{S}({\bs n}_k)/\delta{\bs n_k}$ is the effective magnetic field acting on the impurity with $F_{S}({\bs n}_k)$ being the $k^{\rm th}$ classical spin free energy, and $\delta\bs{B}_k(t)$ is the stochastic magnetic field whose Fourier components $\delta B_{k,\mu}(\omega)$ with $\mu=\bs e_{k,1}, \bs e_{k,2}$ satisfy the fluctuation-dissipation relation \cite{landau2013}:
\begin{equation}
    \langle \delta B_{k,\mu}(\omega)\delta B_{k', \nu}(\omega')\rangle=\underbrace{\frac{\alpha_g\hbar\omega}{\gamma^2 S}\left(\coth{\left(\frac{\hbar\omega}{2k_BT}\right)}-1\right)}_{\displaystyle{S(\omega,T)}}\delta(\omega+\omega')\delta_{\mu\nu}\delta_{k,k'}\,.
\end{equation}

In order to describe the experimental observations \cite{hatter2017,vzitko2018}, we assume $z$ to be an easy-axis (the spin orients perpendicular to the surface), so that the free energy can be written as in the main text:
\begin{align}
    F_{S}({\bs n}_k)=-\frac{\kappa}{2}n_{k,z}^2-\gamma{\bs B}\cdot{\bs n}_k\,,
\end{align}
 where ${\bs B}=B_z{\bs z}$ is  external magnetic field along $z$ and $\kappa$ is the strength of the anisotropy,  which is assumed to be identical for the two spins.  Consequently, the effective magnetic field that determines the dynamics can then be written as ${\bs B}_{k,\rm eff}=B_{k,\rm eff}{\bs z}$, with the magnitude $B_{k,\rm eff}= B_z+(\kappa/\gamma) n_{k,z}$. For the anti-parallel alignment and considering the deterministic direction of the spins to be along the $z$-axis, inserting the  effective field in the LLG equation, we can extract the noise spectrum, for which we find 
 \begin{align}
S_{11}^k&=S_{22}^k=\gamma^2\frac{(\gamma B_{k,\rm eff})^2+(\alpha_g\omega)^2+\omega^2}{[(\gamma B_{k,\rm eff}-\omega)^2+(\alpha_g\omega)^2][(\gamma B_{k,\rm eff}+\omega)^2+(\alpha_g\omega)^2]}S(\omega,T)\,,\nonumber\\
S_{12}^k&=-S_{21}^k=-2i\gamma^2\frac{(-1)^k\omega\gamma B_{k,\rm eff}}{[(\gamma B_{k,\rm eff}-\omega)^2+(\alpha_g\omega)^2][(\gamma B_{k,\rm eff}+\omega)^2+(\alpha_g\omega)^2]}S(\omega,T)\,,
\end{align} 
where $k=1,2$. To give some estimates, we assume $\alpha_1=1.15,~\alpha_2=1.1,~R=2.9,~\alpha_g=0.001$. Considering magnetization anisotropy energy $\kappa=0.1$ meV and $\Omega_0=25$ GHz, we find $T_{2,m}\approx7\,\mu$s allowing around 800 Rabi oscillations to be experimentally observable before the qubit is hampered by the decoherence stemming from magnetic fluctuations. In the presence of a small applied magnetic field, say, $B_z=0.2$ T which corresponds to $\Omega_0=30.5$ GHz, $T_{2,m}\approx 4.5\,\mu$s allowing around 500 Rabi oscillations to be experimentally observable.

\subsection{Decoherence induced by the electron-phonon coupling}

The electron-phonon coupling Hamiltonian can be written as \cite{olivares2014}
\begin{align}
    H_{e-ph}&=\frac{1}{2}g_{ph}\int d\bs r\Psi^\dagger(\bs r)\tau_z \Psi(\bs r) \Phi(\bs r),\\
    \Phi(\bs r)&=\sum_{\bs q}\sqrt{\frac{\hbar\omega_{\bs q}}{2V_0}}(b_{\bs q}e^{i\bs q\cdot\bs r}+b^\dagger_{\bs q}e^{-i\bs q\cdot\bs r})\,,
\end{align}
where $\Psi(\bs r)=(\psi^\dagger_{\uparrow}({\bs r}),\psi^\dagger_{\downarrow}({\bs r}), \psi_{\downarrow}({\bs r}), -\psi_{\uparrow}({\bs r}))^T$ is the electron field operator written in the spin and Nambu basis,  $b_{\bs q}$ ($b_{\bs q}^\dagger$) is  the phonon annihilation (creation) operator with momentum ${\bs q}$, speed velocity $c_s$, and frequency $\omega_{\bs q}=c_sq$ (assuming only acoustic phonons) in the SC of volume $V_0$. This interaction is quantified by the coupling strength  $g_{ph}=\frac{Z\hbar^2\pi^2n_0}{mk_F\sqrt{B}}$, where $Z,~n_0,~B$ are the electron valence from the SC, atomic density and adiabatic bulk modulus, respectively \cite{fetter2012}. The electronic field operator describing the low-energy YSR states can be written as \cite{AkkaravarawongPRR19}:
\begin{equation}
    \Psi(\bs r)\approx\sum_{i=1,2}\phi_{i+}({\bs r})\gamma_i+\phi_{i-}({\bs r})\gamma^\dagger_i\,,
\end{equation}
where $\gamma_{i}$ ($\gamma_{i}^\dagger$) are the annihilation (creation) operators for the in-gap Shiba state at position $i=1,2$, while 
\begin{equation}
    \phi_{i+}(\bs r)=\frac{JS}{\sqrt{N_i}}U_i
    \begin{bmatrix}
    (E_{i+}+\Delta)I_0(\bs r)+I_1(\bs r)\\0\\(E_{i+}+\Delta)I_0(\bs r)-I_1(\bs r)\\0
    \end{bmatrix}\,,
   \end{equation}
and $\phi_{i-}(\bs r)=\tau_y\sigma_yK\phi_{i+}(\bs r)$ are the  eigen-spinors pertaining to energies  $E_{i\pm}=\pm\Delta\left(1-\alpha_i^2\right)/\left(1+\alpha_i^2\right)$, with $K$ being the complex conjugation. Here, $U_i\equiv U(\bs S_i)$ are unitary matrices that align the quantization axis of the Nambu spinor with the direction of impurity spin $i=1,2$, and  $N_i=\frac{(1+\alpha_i^2)^2}{2\pi\nu\Delta\alpha_i}$ is the normalization constant for the $i$th YSR state. The electron-phonon coupling Hamiltonian acting in the low-energy space spanned by the two YSR states can then be written as
    \begin{align}
        H_{e-ph}&=\frac{g_{ph}}{2}\sum_{\bs q}\sqrt{\frac{\hbar\omega_{\bs q}}{2V_0}}(b_{\bs q}+b^\dagger_{-\bs q})[I^o_{ij}({\bs q})\gamma_i^\dagger\gamma_j+I^e_{ij}({\bs q})\gamma_i^\dagger\gamma_j^\dagger+{\rm h. c.}]\,,\\
        I_{ij}^{o,e}({\bs q})&=\int d\bs r~ e^{i\bs q\cdot\bs r} [\phi^\dagger_{1+}(\bs r)\tau_z\phi_{2\pm}(\bs r)-\phi^\dagger_{2\mp}(\bs r)\tau_z\phi_{1-}(\bs r)]\,,
           \end{align}
being overlap integrals between the YSR states and the phonon field. In the limit ${\bs q}\cdot{\bs R}\ll1$, we can approximate  $I_{ij}^{o,e}({\bs q})$ with its ${\bs q}=0$ expression. We have checked numerically that this condition is met in our setup, consequence of the interplay between the phonon energy  $\omega_{q}=c_sq$ which needs to match the qubit splitting, and the inter-impurity distance $R$. Furthermore, the YSR qubit acts in the odd-parity subspace, and thus only the $I_{ij}^{o}(0)$ integrals will be discussed in the following.    
Considering $k_FR\gg 1$,
\begin{align}
    I_{12}^{o}(0)&\approx\frac{2t_h\cos(k_FR+\pi/4)(\alpha_1+\alpha_2)}{(1+\alpha_1^2)(1+\alpha_2^2)\sqrt{\alpha_1\alpha_2}}\frac{R}{\xi}\cos\frac{\theta}{2}\,,
        \label{eq:ep_2}
\end{align}
while we obtain $I_{jj}^{o,e}(0)=0$, which is in itself a novel result (this holds when linearisation of the spectrum around the Fermi level is performed). To evaluate the relaxation, we need to write the above Hamiltonian in the qubit basis. We obtain:
\begin{align}
    H_{q-ph}=\frac{g_{ph}}{2\alpha_1\alpha_2\epsilon_q}I_{12}^o(0)\sum_{\bs q}\sqrt{\frac{\hbar\omega_{\bs q}}{2V_0}}(b_{\bs q}+b^\dagger_{-\bs q})\left(\delta\alpha\,\Sigma_x+t_h\sin(k_FR+\pi/4)\cos\frac{\theta}{2}\,\Sigma_z\right)\,,
\end{align}
which vanishes in the limit $t_h\rightarrow0$, as expected. Moreover, it entails to both population relaxation ($T_{1,ph}$), as well as pure dephasing ($T_{\phi,ph}$). 

The phonon-induced relaxation time $T_{1,ph}$ can be found analogously to $T_{1,m}$ pertaining to the impurities fluctuations calculation. Then, from Eq.~\eqref{eq:ep_2} the relaxation rate can be evaluated as \begin{align}
    \Gamma_{1,ph}=&\frac{2\pi g_{ph}^2(\delta\alpha)^2}{4\hbar\epsilon_q^2\alpha_1^2\alpha_2^2}|I_{12}^o(0)|^2\int \frac{d\bs q}{4\pi^2 x}\hbar\omega_q (2n_q+1)\delta(\epsilon_q-\hbar\omega_{\bs q})
    =\left(\frac{g_{ph}|I_{12}^o(0)|\delta\alpha}{2\hbar c_s\alpha_1\alpha_2}\right)^2\frac{1}{\hbar x}\left(2n(\epsilon_q)+1\right)\,,
\end{align}
where $x$ is the thickness of the 2D SC. We can readily see that  $\Gamma_{1,ph}\propto\cos^2(\theta/2)$, vanishing in the anti-parallel configuration, while becoming maximal in the parallel one. This is in stark contrast to the relaxation induced by impurities fluctuations, $\Gamma_{1,m}\propto\sin^2(\theta/2)$, which vanishes in the parallel configuration. Consequently, the two mechanisms do not compete with each other in the two qubit operation configurations, allowing to separately extract their effects. The reason for such behaviour is that phonons cannot cause spin-flip transitions during the tunneling processes. In the anti-parallel configuration, the tunneling of the YSR quasi-particles involve spin-flips, and thus results in zero coupling. The pure dephasing rate $\Gamma_{\phi,ph}=0$, consequence of the phonon power spectrum $J_{ph}(\omega)\propto\omega^2$ in the current 2D setup  \cite{blum2012}. Then, similarly to magnons, the phonon-induced dephasing entirely originates from longitudinal relaxation, or $T_{2,ph}=2T_{1,ph}$. 

In order to give estimates, let us focus on a 2D Pb SC slab. We assume  $x=10$ nm, i.e. much smaller than the coherence length $\xi$, $c_s\approx10^4$ m/s, and  $g_{ph}\approx1.2\times10^{-8}\sqrt{\mu{\rm eVcm^3}}$. This leads to $\Gamma_{1,ph}(\theta)=1.72\times 10^5\cos^2(\theta/2)$ s$^{-1}$, reaching its maximum at $\theta=0$. The corresponding phonon-induced relaxation time in the parallel configuration is then $T_{1,ph}\approx 5.8 \mu$s, comparable in magnitude to that stemming from impurity fluctuations. 

\section{\label{app:7}Electron-photon coupling in a cavity QED setup}

Next we evaluate the effect of photons (e.g. originating from a microwave cavity coupled to the YSRQ for manipulation and measurement purposes). The electron-photon coupling Hamiltonian reads  \cite{AkkaravarawongPRR19}
\begin{align}
    H_{e-phot}&=\frac{1}{2}\int d\bs r\,\Psi^\dagger(\bs r)[{\bs A}(t)\cdot   \bs{\hat J}+ \bs{\hat J}\cdot{\bs A}(t)]\Psi(\bs r)\approx\sum_{i,j=1,2} M^{o}_{ij}(t)\gamma_i^\dagger\gamma_j+M^{e}_{ij}(t)\gamma_i^\dagger\gamma_j^\dagger+h.c.\,,\nn\\
    M^{e,o}_{ij}(t)&=\int d\bs r \bs A(t)\cdot\bs J(\bs r)\,,\\
    {\bs J}^{e,o}_{ij}(\bs r)&=\frac{e\hbar}{2mi}[\phi^\dagger_{i\pm}({\bs r})\nabla\phi_{j+}(\bs r)-\phi^\dagger_{j\mp}(\bs r)\nabla\phi_{i-}(\bs r)]\,,
    \end{align}
where the $\bs{\hat J}$ is the current operator that couples to the  vector potential ${\bs A}(t)$ of the electromagnetic field via the substitution $-i\hbar\nabla\rightarrow -i\hbar\nabla+e{\bs A}(t)$. For simplicity, we assume ${\bs A}(t)$  constant in space over the size of the YSRQ, since the wavelength of the photons resonant with $\epsilon_q$ are  longer than the coherence length. We mention tht the diagonal terms $M_{jj}^{o}(t)=0$, since the localized states do not carry any current. Using $\bs{\mathcal{E}}(t)=-\partial{\bs A}(t)/\partial t$, with $\bs{\mathcal{E}}(t)$ being the electric field, allows to write for the odd-parity sector term in the Fourier space:
\begin{align}
    |M^{o}_{12}(\omega)|\approx&\frac{e\Delta\xi\,\bs {\mathcal{E}}(\omega)\cdot\hat{\bs R}}{\hbar\omega }\frac{t_h\cos(\theta/2)\cos(k_FR+\pi/4)}{\sqrt{\alpha_1\alpha_2}(1+\alpha_1^2)(1+\alpha_2^2)}\left[1+\alpha_1\alpha_2+\frac{R}{\xi}(\alpha_1\alpha_2-1)\right]\,,
    \label{eq:ep_3}
\end{align}
where we utilised $i\omega {\bs A}(\omega)=\bs{\mathcal E}(\omega)$ and $\hat{\bs R}={\bs R}/R$. To give estimates for the coupling strength, we note that in microwave cavities with frequencies comparable to the qubit splitting the electric field can be as large as  $\mathcal{E}\approx0.2$ V/m \cite{blais2021} which, when using the same YSRQ parameters as in the previous section, leads to a coupling strength $|M_{12}^o(\epsilon_q)|/\hbar\approx 10 $ MHz in the parallel alignment of the magnetic impurities.

\end{document}